\documentclass[12pt]{iopart}
\usepackage[utf8]{inputenc}

\usepackage{harvard}
\usepackage{graphicx}
\usepackage{appendix}
\usepackage{caption} 
\usepackage{xcolor}
\usepackage{courier}
\begin{document}

\title[DaRT finite-element modeling]{Finite-element modeling of the alpha particle dose of realistic sources used in Diffusing Alpha-emitters Radiation Therapy}

\author{G. Heger, L. Arazi\footnote{Corresponding Author.}}

\address{Unit of Nuclear Engineering, Faculty of Engineering Sciences, Ben-Gurion University of the Negav, P.O.B. 653 Be'er-Sheva 8410501, Israel}
\ead{larazi@bgu.ac.il}

\vspace{10pt}
\begin{indented}
\item[]October 2021
\end{indented}

\begin{abstract}
    Diffusing Alpha-emitters Radiation Therapy (DaRT) is a new method for treating solid tumors using alpha particles. Unlike conventional radiotherapy, where the physical models for dose calculations are known and routinely used, for DaRT a new framework, called the Diffusion-Leakage (DL) model, had to be developed. In this work we provide a detailed description of a finite-element numerical scheme for solving the time-dependent DL model equations for cylindrical DaRT sources (``seeds'') of finite diameter and length in two dimensions. Using a fully-implicit scheme and adaptive time step, this approach allows to accurately follow temporal transients ranging from seconds to days. In addition to the full two-dimensional calculation, we further provide a closed-form approximation and a simple one-dimensional scheme to solve the equations for infinitely-long cylindrical sources. These simpler solutions can be used both to validate the two-dimensional code and to enable efficient parameter scans to study the properties of DaRT seed lattices. We compare these approximations to the full 2D solution over the relevant parameter space, providing guidelines on their usability and limitations. 
\end{abstract}

%
%
\submitto{\PMB}
%
%
%

\section{Introduction}

\begin{figure}
	\begin{center}
		\includegraphics{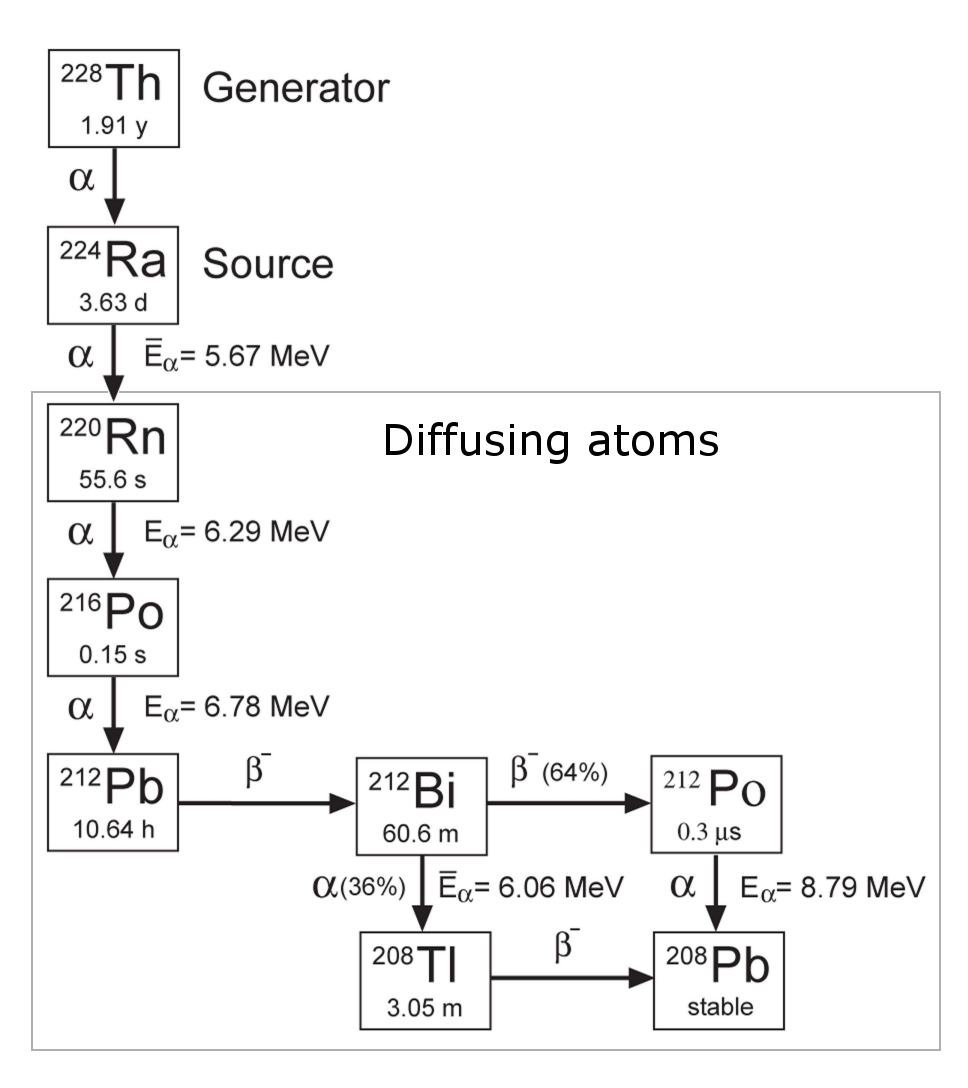}
		\caption{The $^{224}$Ra decay chain. Data taken from the NuDat2 database website, https://www.nndc.bnl.gov/nudat2/} \label{fig:Ra224_decay_chain}
	\end{center}
\end{figure}

Diffusing Alpha-Emitters Radiation Therapy (DaRT), first introduced by \citeasnoun{Arazi2007}, is a new form of radiation therapy, which enables the treatment of solid tumors using alpha particles. In DaRT, tumors are treated with sources (``seeds'') carrying low activities of $^{224}$Ra below their surface. Once inside the tumor, the seeds continuously release from their surface the short-lived daughter atoms of $^{224}$Ra: $^{220}$Rn, $^{216}$Po, $^{212}$Pb, $^{212}$Bi, $^{212}$Po and $^{208}$Tl. These spread by diffusion (with possible contribution by convective effects), creating -- primarily through their alpha decays -- a ``kill-region'' measuring several mm in diameter around each seed. The full decay chain, along with the isotopes' half-lives, decay modes and mean alpha particle energies is shown in figure \ref{fig:Ra224_decay_chain}.

\citationmode{abbr}

DaRT was, and still is, extensively investigated in {\it in vitro} and {\it in vivo} preclinical studies on a large number of cancer types,  as a stand-alone treatment \cite{Arazi2007,Cooks2008,Cooks2009a,Lazarov2011,Cooks2012}, in combination with chemotherapy \cite{Cooks2009b,Horev-Drori2012,Milrot2013,Reitkopf-Brodutch2015}, and as a stimulator of local and systemic anti-tumor immune response \cite{Keisari2014,Confino2015,Confino2016,Domankevich2019,Domankevich2020,Keisari2020,Keisari2021}. Since 2017, DaRT is under clinical investigations in human patients, starting with locally advanced and recurrent squamous cell carcinoma of the skin and head and neck \cite{Popovtzer2019}. Results of the first-in-man trial were highly promising in terms of both efficacy and safety: all treated tumors shrank drastically (by $\sim30-100\%$), beginning in the first few days after the treatment, with $\sim80\%$ of the tumors exhibiting complete response. Adverse affects were mild to moderate, with no observable local or systemic radiation-induced damage to healthy tissue. The alpha dose to all organs, resulting from $^{212}$Pb leaving the tumor through the blood, was calculated to be on the cGy level, with blood and urine activity measurements consistent with the prediction of a biokinetic model \cite{Arazi2010}. In one patient, untreated lesions shrank and disappeared when another was treated with DaRT (in the absence of any other treatment), suggesting an abscopal effect \cite{Bellia2019}.

An in-depth description of the underlying physics of DaRT was given in \cite{Arazi2020}. As a first step towards treatment planning, it introduced a simplified theoretical approach: the ``Diffusion-Leakage (DL) model''. Its underlying assumptions are that the medium into which the daughters of $^{224}$Ra are released is homogeneous, isotropic and does not change with time, and that convective effects have a short correlation length and can therefore be described by an effective isotropic diffusion term. A further assumption is that $^{212}$Pb and $^{212}$Bi can be cleared from the tumor through the blood at a uniform local rate. The discussion in \cite{Arazi2020} showed that of the six radionuclides released from the source, one only needs to model the migration of $^{220}$Rn, $^{212}$Pb and $^{212}$Bi as their respective short-lived daughters are in local secular equilibrium. 

The focus of \cite{Arazi2020} was the alpha particle dose of the ideal point source, which constitutes the basic building block for any arbitrary configuration of line sources. In this work, we expand the treatment to realistic cylindrical seeds of finite diameter and length. We derive closed-form asymptotic and approximate time-dependent solutions to the diffusion-leakage model for infinite cylindrical sources, and describe in detail numerical schemes for solving the complete time-dependent problem in both one and two dimensions, to model the alpha particle dose for realistic seed geometries. We compare the 2D numerical solution to the 1D case and to other approximations, and provide examples for calculating the alpha particle dose in hexagonal seed lattices. The presented calculations can serve for generating dose lookup tables as a pragmatic approach to treatment planning.

\section{The diffusion-leakage model in cylindrical coordinates}

The underlying assumptions of the DL model, described at length in \cite{Arazi2020}, are:

\begin{itemize}
	\item The migration of atoms inside the tumor is governed by diffusion.
	\item The tissue is homogeneous, isotropic, and time-independent, and thus the diffusion coefficients are constant.
	\item It is sufficient to model the migration of $^{220}$Rn, $^{212}$Pb and $^{212}$Bi. $^{216}$Po, $^{212}$Po and $^{208}$Tl are in local secular equilibrium with their respective parent isotopes (with suitable branching ratios for the latter two).
	\item $^{212}$Pb migration can be described using a single effective diffusion coefficient.
	\item $^{212}$Pb atoms reaching major blood vessels are trapped in red blood cells and are then immediately cleared from the tumor. This is represented by a single constant sink term.
	\item $^{220}$Rn atoms do not form chemical bonds and are very short-lived, and therefore the equation for Rn does not include a sink term.
	\item The equation for $^{212}$Bi does contain a sink term, but it is considered a second-order effect and generally set to zero.
\end{itemize}

\noindent {We consider the case of a cylindrical source of radius $R_0$ and length $l$ along the $z$ axis, and assume axial symmetry. Under the above assumptions, in cylindrical coordinates $(r,z)$ the equations describing the dynamics of the main daughter atoms in the decay chain -- $^{220}$Rn, $^{212}$Pb and $^{212}$Bi -- are:}

\begin{equation} \label{eq:Rn_diffusionEqn_2D}
	\fl \qquad \qquad \frac{\partial n_{Rn}}{\partial t}=D_{Rn}\left(\frac{1}{r}\frac{\partial}{\partial r}\left(r\frac{\partial n_{Rn}}{\partial r}\right)+\frac{\partial^2 n_{Rn}}{\partial z^2} \right)-\lambda_{Rn}n_{Rn}
\end{equation}

\begin{equation} \label{eq:Pb_diffusionEqn_2D}
	\fl \qquad \qquad \frac{\partial n_{Pb}}{\partial t}=D_{Pb}\left(\frac{1}{r}\frac{\partial}{\partial r}\left(r\frac{\partial n_{Pb}}{\partial r}\right)   +\frac{\partial^2 n_{Pb}}{\partial z^2} \right)+\lambda_{Rn}n_{Rn}-\left(\lambda_{Pb}+\alpha_{Pb}\right)n_{Pb}
\end{equation}

\begin{equation} \label{eq:Bi_diffusionEqn_2D}
	\fl \qquad \qquad \frac{\partial n_{Bi}}{\partial t}=D_{Bi}\left(\frac{1}{r}\frac{\partial}{\partial r}\left(r\frac{\partial n_{Bi}}{\partial r}\right)   +\frac{\partial^2 n_{Bi}}{\partial z^2} \right)+\lambda_{Pb}n_{Pb}-\left(\lambda_{Bi}+\alpha_{Bi}\right)n_{Bi}
\end{equation}
where  $n_{Rn}$, $n_{Pb}$, $n_{Bi}$, $D_{Rn}$, $D_{Pb}$, $D_{Bi}$ and $\lambda_{Rn}$, $\lambda_{Pb}$, $\lambda_{Bi}$ are the number densities, diffusion coefficients and decay rate constants of $^{220}$Rn, $^{212}$Pb and $^{212}$Bi, respectively. $\alpha_{Pb}$ and $\alpha_{Bi}$ are the leakage rate constants, accounting for clearance through the blood, of $^{212}$Pb and $^{212}$Bi. As discussed in \cite{Arazi2020}, one can generally assume $\alpha_{Bi}=0$, as will be done here. The boundary conditions, for $r\rightarrow R_0$ and $|z|\leq l/2$ ($z=0$ at the seed mid plane), are:

\begin{equation}
    \lim_{r\rightarrow R_0} 2\pi r j_{Rn}\left(r,z,t\right)=P_{des}(Rn)\frac{\Gamma_{Ra}^{src}(0)}{l} e^{-\lambda_{Ra}t}   \label{eq:Rn_bnd_cond_src}
\end{equation}

\begin{equation}
     \lim_{r\rightarrow R_0} 2\pi r j_{Pb}(r,z,t)=\left(P_{des}^{eff}(Pb)-P_{des}(Rn)\right)\frac{\Gamma_{Ra}^{src}(0)}{l} e^{-\lambda_{Ra}t} \label{eq:Pb_bnd_cond_src}
\end{equation}

\begin{equation}
     \lim_{r\rightarrow R_0} 2\pi r j_{Bi}(r,z,t)=0 \label{eq:Bi_bnd_cond_src}
\end{equation}
where $j_x=-D_x\partial{n_x}/\partial{r}$, with $x$ representing $^{220}$Rn, $^{212}$Pb and $^{212}$Bi. $\Gamma_{Ra}^{src}(0)$ is the initial $^{224}$Ra activity on the source (assumed to be uniform) and $\lambda_{Ra}$ is the $^{224}$Ra decay rate constant. $P_{des}(Rn)$ and $P_{des}^{eff}(Pb)$ are the desorption probabilities of $^{220}$Rn and $^{212}$Pb, respectively, representing the probability that a decay of a $^{224}$Ra on the source will lead to the emission of either a $^{220}$Rn or $^{212}$Pb atom into the tumor; we use the term ``effective'' for $P_{des}^{eff}(Pb)$, because it includes several emission pathways, as discussed in \cite{Arazi2020}. For $|z|>l/2$, $\lim_{r\rightarrow R_0} r j_x(r,z,t) = 0$ for the three isotopes.

The solution for equations (\ref{eq:Rn_diffusionEqn_2D}-\ref{eq:Bi_diffusionEqn_2D}) provides the number densities $n_{Rn}(r,z,t)$, $n_{Pb}(r,z,t)$ and $n_{Bi}(r,z,t)$. The alpha dose is calculated under the assumption that the range of alpha particles is much smaller than the dominant diffusion length of the problem (see below). The dose from source insertion to time $t$ is comprised of two contributions: one arising from the summed alpha particle energy of the pair $^{220}$Rn+$^{216}$Po, and the other from the the alpha decay of either $^{212}$Bi or $^{212}$Po:

\begin{equation}
    Dose_{\alpha}(RnPo;\,r,z,t)=\frac{E_{\alpha}(RnPo)}{\rho}\int_{0}^{t}\lambda_{Rn}n_{Rn}(r,z,t')dt' \label{eq:Dose_RnPo(r,t) general form}
\end{equation}

\begin{equation}
    Dose_{\alpha}(BiPo;\,r,z,t)=\frac{E_{\alpha}(BiPo)}{\rho}\int_{0}^{t}\lambda_{Bi}n_{Bi}(r,z,t')dt' \label{eq:Dose_BiPo(r,t) general form}
\end{equation}
where $E_{\alpha}(RnPo)=(6.288+6.778)\:\textrm{MeV}=13.066\:\textrm{MeV}$ is the total alpha particle energy of $^{220}$Rn and $^{216}$Po, $E_{\alpha}(BiPo)=7.804\:\textrm{MeV}$ is the weighted-average energy of the alpha particles emitted by $^{212}$Bi and $^{212}$Po, and $\rho$ is the tissue density. In what follows, we define the ``asymptotic dose'' as the dose delivered from source insertion to infinity -- in practice, over several weeks.

The discussion in \cite{Arazi2020} has shown the the spread of $^{220}$Rn, $^{212}$Pb and $^{212}$Bi is governed by their respective {\it diffusion lengths}, defined as:

\begin{equation}
    L_{Rn}=\sqrt{\frac{D_{Rn}}{\lambda_{Rn}-\lambda_{Ra}}} \label{eq:LRn}
\end{equation}

\begin{equation}
    L_{Pb}=\sqrt{\frac{D_{Pb}}{\lambda_{Pb}+\alpha_{Pb}-\lambda_{Ra}}} \label{eq:LPb}
\end{equation}

\begin{equation}
    L_{Bi}=\sqrt{\frac{D_{Bi}}{\lambda_{Bi}+\alpha_{Bi}-\lambda_{Ra}}} \label{eq:LBi}
\end{equation}

The diffusion length is a gross measure of the average displacement of an atom from its creation site, to the point of its decay  or clearance by the blood. For a point source the radial dependence of the number densities and alpha dose components comprises terms proportional to $e^{-r/L_x}/(r/L_x)$. As noted in \cite{Arazi2020} typical ranges are $L_{Rn}\sim0.2-0.4$\;mm, $L_{Pb}\sim0.3-0.7$\;mm and $L_{Bi}/L_{Pb}\sim0.1-0.2$.

Another important parameter introduced in \cite{Arazi2020} is the $^{212}$Pb {\it leakage probability} $P_{leak}(Pb)$, defined as the probability that a $^{212}$Pb atom released from the source is cleared from the tumor by the blood before its decay. The leakage probability therefore reflects the competition between $^{212}$Pb radioactive decay and clearance through the blood, such that:

\begin{equation}
    P_{leak}(Pb)=\frac{\alpha_{Pb}}{\lambda_{Pb}+\alpha_{Pb}}
\end{equation}
The typical range of values is $P_{leak}(Pb)\sim0.2-0.8$.

\section{Asymptotic and approximate time-dependent solutions for an infinitely long cylindrical source}

As discussed in \cite{Arazi2020}, at long times after source insertion into the tumor the number densities reach an asymptotic form: $n_x^{asy}(r,z,t)=\widetilde{n}_x(r,z)e^{-\lambda_{Ra}t}$. For $^{220}$Rn this condition is satisfied within several minutes throughout the tumor, while for $^{212}$Pb and $^{212}$Bi the asymptotic form is attained within a few days, depending on the distance from the source. 

In \ref{Appendix A} we provide a derivation of the closed-form asymptotic solution of the DL model equations for an infinitely long cylindrical source. For $^{220}$Rn we get:

\begin{equation} \label{eq:nRn_asy}
    n_{Rn}^{asy}(r,t)=A_{Rn}\;K_0\left(\frac{r}{L_{Rn}}\right) e^{-\lambda_{Ra}t} 
\end{equation}
where:

\begin{equation}
    A_{Rn}=\frac{P_{des}(Rn)\left(\Gamma_{Ra}^{src}(0)/l\right)}{2\pi D_{Rn} \, (R_0/L_{Rn}) K_1\left(R_0/L_{Rn}\right)} \label{eq:ARn}
\end{equation}

\noindent{In these expressions $K_0(\xi)$ and $K_1(\xi)$ are modified Bessel functions of the second kind:}

\begin{eqnarray}
    K_{0}\left(\xi\right)=\int_{0}^{\infty}\frac{\cos\left(\xi t\right)}{\sqrt{t^{2}+1}}dt \\
    K_{1}(\xi)=-\frac{dK_{0}}{d\xi}
\end{eqnarray}

$K_0(r/L)$ is a steeply-falling function, and is the cylindrical analogue to $\exp(-r/L)/(r/L)$ appearing in expressions for the number densities and dose of the point source \cite{Arazi2020}. As shown in \ref{Appendix A}, for $^{212}$Pb we have:

\begin{eqnarray} \label{eq:nPb_asy}
    n_{Pb}^{asy}(r,t)=\left(A_{Pb}\;K_{0}\left(\frac{r}{L_{Rn}}\right)+B_{Pb}\;K_{0}\left(\frac{r}{L_{Pb}}\right)\right) e^{-\lambda_{Ra}t} 
\end{eqnarray}
where:

\begin{eqnarray}
    \fl \qquad \qquad A_{Pb}=\frac{L_{Rn}^{2}L_{Pb}^{2}}{L_{Rn}^{2}-L_{Pb}^{2}}\frac{\lambda_{Rn}}{D_{Pb}}A_{Rn} \\ \label{eq:APb}
	\fl \qquad \qquad B_{Pb} = \frac{\left(P_{des}^{eff}(Pb)-P_{des}(Rn)\right)\left(\Gamma_{Ra}^{src}(0)/l\right)}{ 2\pi D_{Pb} \, (R_0/L_{Pb})K_1(R_0/L_{Pb}) } - 
	A_{Pb}\frac{ (R_0/L_{Rn})K_1(R_0/L_{Rn}) }{ (R_0/L_{Pb})K_1(R_0/L_{Pb}) } \label{eq:BPb}
\end{eqnarray}
Finally, for $^{212}$Bi we get:

\begin{equation}
    \fl \qquad \qquad n_{Bi}^{asy}(r,t) = \left(A_{Bi}\;K_{0}\left(\frac{r}{L_{Rn}}\right)+B_{Bi}\;K_{0}\left(\frac{r}{L_{Pb}}\right)+C_{Bi}\;K_{0}\left(\frac{r}{L_{Bi}}\right)\right) e^{-\lambda_{Ra}t} \label{eq:nBi_asy}    
\end{equation}
where:

\begin{equation}
    A_{Bi}=\frac{L_{Rn}^{2}L_{Bi}^{2}}{L_{Rn}^{2}-L_{Bi}^{2}}\frac{\lambda_{Pb}}{D_{Bi}}A_{Pb} \label{eq:ABi}
\end{equation}

\begin{equation}
    B_{Bi}=\frac{L_{Pb}^{2}L_{Bi}^{2}}{L_{Pb}^{2}-L_{Bi}^{2}}\frac{\lambda_{Pb}}{D_{Bi}}B_{Pb} \label{eq:BBi}
\end{equation}

\begin{equation}
    C_{Bi}=- \frac{ (R_0/L_{Rn})K_1(R_0/L_{Rn})A_{Bi} + (R_0/L_{Pb})K_1(R_0/L_{Pb})B_{Bi} }{ (R_0/L_{Bi})K_1(R_0/L_{Bi}) } \label{eq:CBi}  
\end{equation}
As shown in \ref{Appendix A} the expressions above can also describe the limit of an infinite line source in the limit $R_0/L_x\rightarrow 0$.

To approximately account for the buildup stage of the solution, one can assume that it is uniform throughout the tumor, i.e., independent of the distance from the source. Under this ``0D'' temporal approximation, discussed in \cite{Arazi2020} for a point source and adapted here for the cylindrical case, one can write:

\begin{equation}
    n_{Rn}^{0D}(r,t)=A_{Rn}\;K_0\left(\frac{r}{L_{Rn}}\right) \left( e^{-\lambda_{Ra}t} - e^{-\lambda_{Rn}t} \right) \label{eq:nRn_0D}
\end{equation}
and
\begin{eqnarray}
    \fl \qquad n_{Bi}^{0D}(r,t) = \nonumber \\ \fl \qquad  \left(A_{Bi}\;K_{0}\left(\frac{r}{L_{Rn}}\right)+B_{Bi}\;K_{0}\left(\frac{r}{L_{Pb}}\right)+C_{Bi}\;K_{0}\left(\frac{r}{L_{Bi}}\right)\right) \left( e^{-\lambda_{Ra}t} - e^{-(\lambda_{Pb}+\alpha_{Pb})t} \right) \nonumber \\ \label{eq:nBi_0D}
\end{eqnarray}

\noindent{Under this approximation, the asymptotic alpha dose components are:}

\begin{equation} \label{eq:Dose(RnPo)_asy_0D}
    \fl \qquad Dose_{\alpha}^{asy}(RnPo;\,r)=\frac{E_{\alpha}(RnPo)}{\rho}\lambda_{Rn}A_{Rn}\;K_0\left(\frac{r}{L_{Rn}}\right) \left(\tau_{Ra}-\tau_{Rn}\right)
\end{equation}

\begin{eqnarray} \label{eq:Dose(BiPo)_asy_0D}
    \fl \qquad Dose_{\alpha}^{asy}(BiPo;\,r)= \nonumber \\ \fl \qquad \frac{E_{\alpha}(BiPo)}{\rho}\lambda_{Bi}\left(A_{Bi}\;K_{0}\left(\frac{r}{L_{Rn}}\right)+B_{Bi}\;K_{0}\left(\frac{r}{L_{Pb}}\right)+C_{Bi}\;K_{0}\left(\frac{r}{L_{Bi}}\right)\right) \left(\tau_{Ra}-\tau_{Pb}^{eff}\right)  \nonumber \\
\end{eqnarray}
where $\tau_{Ra}=1/\lambda_{Ra}$, $\tau_{Rn}=1/\lambda_{Rn}$ and $\tau_{Pb}^{eff}=1/(\lambda_{Pb}+\alpha_{Pb})$. The error introduced by the 0D approximation is of the order of the ratio between mean lifetimes of $^{220}$Rn and $^{212}$Pb and that of $^{224}$Ra, i.e. $\tau_{Rn}/\tau_{Ra}\sim10^{-4}$ and $\tau_{Pb}/\tau_{Ra}\sim0.1$, respectively. 

\section{Finite-element one-dimensional time-dependent solution for an infinite cylindrical source}

\subsection{Numerical scheme}

A complete time-dependent solution to the DL model can be done numerically using a finite-element approach. For the one-dimensional case, i.e., infinite cylindrical or line source along the $z$ axis, we solve equations (\ref{eq:Rn_diffusionEqn_2D}-\ref{eq:Bi_diffusionEqn_2D}) setting $\partial^2 n_x/\partial z^2=0$. The solution therefore depends solely on the radial coordinate $r$. We divide our domain into concentric cylindrical shells, enumerated $i=1...N_r$, of equal radial width $\Delta r$. The radius of the source is $R_0$; we choose $\Delta r$ such that $R_0/\Delta r$ is an integer number, and $\Delta r$ is considerably smaller than $L_{Rn}$ and $L_{Pb}$ ($L_{Bi}$ has a negligible effect on the solution and therefore should not constrain $\Delta r$). The central radius of the $i$-th shell is:

\begin{equation} \label{eq:ri_1D}
    r_i=R_0+(i-\frac{1}{2})\Delta r    
\end{equation}

We employ a fully implicit scheme \cite{Numerical_Recipes}, thereby assuring the solution is stable. The time steps are changed adaptively according to the relative change in the solution between the current step and the previous one, as explained below. 

The average number densities in the $i$-th shell are $n_{Rn,i}$, $n_{Pb,i}$ and $n_{Bi,i}$. We enumerate the time steps by $p$. For shells away from the source surface, with $1<i\leq N_r$, the DL model equations take the discrete implicit form:

\begin{eqnarray} \label{eq:general_discrete_1D_eq}
\fl \qquad \frac{n^{(p+1)}_{x,i}-n^{(p)}_{x,i}}{\Delta t}= \nonumber \\
\fl \qquad D_{x}\bigg(\frac{n^{(p+1)}_{x,i+1}+n^{(p+1)}_{x,i-1}-2n^{(p+1)}_{x,i}}{\Delta r^2} +\frac{1}{r_i}\frac{n^{(p+1)}_{x,i+1}-n^{(p+1)}_{x,i-1}}{2\Delta r}\bigg)-(\lambda_{x}+\alpha_{x})n^{(p+1)}_{x,i}+s^{(p+1)}_{x,i} \nonumber \\ 
\end{eqnarray}
where $x$ stands for Rn, Pb and Bi and $\alpha_{Rn}=0$. Outside our domain we set the number densities to zero, such that in eq. (\ref{eq:general_discrete_1D_eq}) for $i=N_r$ $n_{x,i+1}=0$.

As shown in \ref{Appendix B}, for the $i=1$ shell, immediately outside of the source, using the boundary conditions, eq. (\ref{eq:Rn_bnd_cond_src})-(\ref{eq:Bi_bnd_cond_src}), gives:

\begin{eqnarray} \label{eq:1D_eq_first_shell}
\fl \qquad \frac{n^{(p+1)}_{x,1}-n^{(p)}_{x,1}}{\Delta t}= \frac{D_x}{\Delta r^2}\left(\frac{1+\Delta r/R_0}{1+\Delta r/2R_0}\right)\left(n^{(p+1)}_{x,2}-n^{(p+1)}_{x,1}\right)-(\lambda_{x}+\alpha_{x})n^{(p+1)}_{x,1}+s^{(p+1)}_{x,1} \nonumber \\ 
\end{eqnarray}

\noindent{The source terms $s^{p+1}_{x,i}$ appearing in eq. (\ref{eq:general_discrete_1D_eq}) and (\ref{eq:1D_eq_first_shell}) are:} 

\begin{eqnarray} \label{eq:1D_source_terms}
	\fl \qquad s_{Rn,i}^{p+1} = \frac{P_{des}(Rn)\left(\Gamma_{Ra}^{src}(0)/l\right) e^{-\lambda_{Ra}t_{p+1}} }{2\pi R_0\Delta r\left(1+\Delta r/2R_0\right) } \delta_{i,1} \\
	\fl \qquad s_{Pb,i}^{p+1} = \frac{\left(P_{des}^{eff}(Pb)-P_{des}(Rn)\right)\left(\Gamma_{Ra}^{src}(0)/l\right) e^{-\lambda_{Ra}t_{p+1}} }{2\pi R_0\Delta r\left(1+\Delta r/2R_0\right)} \delta_{i,1} + \lambda_{Rn}n_{Rn,i}^{p+1} \\
	\fl \qquad s_{Bi,i}^{p+1} = \lambda_{Pb}n_{Pb,i}^{p+1}
\end{eqnarray}
where $\delta_{i,1}=1$ for $i=1$ and zero otherwise. Rearranging eq. (\ref{eq:general_discrete_1D_eq}) and (\ref{eq:1D_eq_first_shell}), we get the general form:

\begin{equation}
    n^{(p)}_{x,i} + s^{(p+1)}_{x,i}\Delta t = A_{i,i-1}^{(x)}n^{(p+1)}_{x,i-1} + A_{i,i}^{(x)}n^{(p+1)}_{x,i} + A_{i,i+1}^{(x)}n^{(p+1)}_{x,i+1} \label{eq:general_1D_eq_with_A_ij}
\end{equation}

The matrix coefficients introduced in eq. (\ref{eq:general_1D_eq_with_A_ij}) depend on the value of $i$, reflecting the boundary conditions for $i=1$ and $i=N_r$. Retaining terms up to first order in $\Delta r/r_i$ the different cases are summarized below:

\begin{eqnarray}
\fl \qquad \qquad A_{i,i-1}^{(x)} & = -\frac{D_x\Delta t}{\Delta r^2} \left(1-\frac{\Delta r}{2r_i}\right)        & 1<i \leq N_r \\
\fl \qquad \qquad A_{i,i}^{(x)}   & = 1+\frac{D_x\Delta t}{\Delta r^2}\left(1+\frac{\Delta r}{2r_i}\right)+\left(\lambda_x+\alpha_x\right)\Delta t \qquad & i=1 \nonumber \\
                                  & = 1+\frac{2D_x\Delta t}{\Delta r^2}+\left( \lambda_x+\alpha_x \right)\Delta t & 1<i \leq N_r \\
\fl \qquad \qquad A_{i,i+1}^{(x)} & = -\frac{D_x\Delta t}{\Delta r^2} \left(1+\frac{\Delta r}{2r_i}\right)        & 1\leq i < N_r 
\end{eqnarray}

\noindent{with $r_i$ given in eq. (\ref{eq:ri_1D}). Writing eq. (\ref{eq:general_1D_eq_with_A_ij}) in matrix form:}

\begin{eqnarray} \label{matrixEq_1D}
\fl
\resizebox{1\hsize}{!}{$
    \left( \matrix{
    n^{(p)}_{x,1} \cr
    n^{(p)}_{x,2} \cr
    \vdots \cr
    n^{(p)}_{x,i} \cr
    \vdots\cr
    \cr
    n^{(p)}_{x,N_r-1} \cr
    n^{(p)}_{x,N_r} \cr
} \right)
+
\left( \matrix{
    s^{(p+1)}_{x,1} \cr
	s^{(p+1)}_{x,2} \cr
	\vdots \cr
	s^{(p+1)}_{x,i} \cr
	\vdots\cr
	\cr
	s^{(p+1)}_{x,N_r-1} \cr
	s^{(p+1)}_{x,N_r} \cr
} \right)\Delta t_{p+1}
=
\left( \matrix{
    A_{1,1}^{(x)}  & A_{1,2}^{(x)} & 0               & \cdots        &                       &                       &  0\cr
    A_{2,1}^{(x)}  & A_{2,2}^{(x)} & A_{2,3}^{(x)}   & 0             &\cdots                 &                       &  0\cr
                   & \ddots        & \ddots          & \ddots        &                       &                       &  &\cr
    \vdots         & 0             & A_{i,i-1}^{(x)} & A_{i,i}^{(x)} & A_{i,i+1}^{(x)}       & 0                     & \vdots\cr
                   &               &                 & \ddots        & \ddots                & \ddots                &\cr
                   &               &                 &               &                       &                       &\cr
    0              & \cdots        &                 & 0             & A_{N_r-1,N_r-2}^{(x)} & A_{N_r-1,N_r-1}^{(x)} & A_{N_r-1,N_r}^{(x)} \cr
    0              &               &                 & \cdot         & 0                     & A_{N_r,N_r-1}^{(x)}   & A_{N_r,N_r}^{(x)}
} \right)
\left( \matrix{
    n^{(p+1)}_{x,1} \cr
    n^{(p+1)}_{x,2} \cr
    \vdots \cr
    n^{(p+1)}_{x,i} \cr
    \vdots\cr
    \cr
    n^{(p+1)}_{x,N_r-1} \cr
    n^{(p+1)}_{x,N_r} \cr
    } \right)
    $} 
    \nonumber \\ \nonumber \\
\end{eqnarray}
which can be written as $\mathbf{n}^{(p)}_x+\mathbf{s}^{(p+1)}_x\Delta t=\mathbf{A}_x \mathbf{n}^{(p+1)}_x$. Multiplying on the left by the inverse of $\mathbf{A}_x$, we get:

\begin{equation} \label{eq:1D matrix solution}
    \mathbf{n}^{(p+1)}_x=\mathbf{A}_x^{-1}(\mathbf{n}^{(p)}_x+\mathbf{s}^{(p+1)}_x\Delta t),     
\end{equation}
which completes the solution for the $p+1$ step. Note that although the source terms are calculated for the $p+1$ step they are, in fact, known when the matrix equations are solved. The reason is that in the $p+1$ step we first solve for $^{220}$Rn, then for $^{212}$Pb and lastly for $^{212}$Bi. The source term for $^{220}$Rn depends only on time, those of $^{212}$Pb are found using the $p+1$ solution for $^{220}$Rn, and those of $^{212}$Bi -- using the $p+1$ solution for $^{212}$Pb. Another point to take into account is that since $\Delta t$ is changed along the calculation and the matrix coefficients depend on $\Delta t$, they must be updated accordingly in each step.

\bigskip
\noindent{The alpha dose components are also updated in each step:}

\begin{equation}
    \fl \qquad \qquad Dose_{\alpha}^{(p+1)}(RnPo;\,i)=Dose_{\alpha}^{(p)}(RnPo;\,i)+\frac{E_{\alpha}(RnPo)}{\rho}\lambda_{Rn} n_{Rn,i}^{(p+1)}\Delta t    
\end{equation}

\begin{equation}
    \fl \qquad \qquad Dose_{\alpha}^{(p+1)}(BiPo;\,i)=Dose_{\alpha}^{(p)}(BiPo;\,i)+\frac{E_{\alpha}(BiPo)}{\rho}\lambda_{Bi} n_{Bi,i}^{(p+1)}\Delta t    
\end{equation}

At the end of the $p+1$ step, $\Delta t$ is updated based on the relative change in the solution. This can be done in a number of ways. A particular choice, implemented here, was to consider the relative change in the total dose (sum of the $RnPo$ and $BiPo$ contributions) at a particular point of interest $r_{i_0}$ (e.g., at 2 mm):

\begin{equation} \label{eq:DART1D_dt_change}
    \fl \qquad \qquad \Delta t_{new} = {\Delta t} \cdot \frac{\epsilon_{tol}}{\left(Dose_{\alpha}^{(p+1)}(tot;i_0)-Dose_{\alpha}^{(p)}(tot;i_0)\right)/Dose_{\alpha}^{(p)}(tot;i_0)} 
\end{equation}
where $\epsilon_{tol}$ is a preset tolerance parameter. For practicality, one can further set upper and lower limits on $\Delta t$ to balance between calculation time and accuracy. Although the initial time step should be small compared to $^{220}$Rn half-life, its particular value has little effect on the accuracy of the calculated asymptotic dose. 

\subsection{DART1D: tests and comparisons}

The finite-element scheme described above was implemented in MATLAB in a code named ``DART1D''. The solution of eq. (\ref{eq:1D matrix solution}), the critical part of the calculation, was done using a tridiagonal matrix solver employing the Thomas algorithm \cite{MATLAB_tridiagonal}. The use of this solver was found to be $\sim4$ times faster than MATLAB's mldivide (`\textbackslash') tool, which was, in turn, about 3-fold faster than inverting the matrix using \texttt{inv(A)}. The code was found to converge to sub-percent level for a modest choice of the discretization parameter values. For example, setting $\epsilon_{tol}=10^{-2}$, $\Delta r=0.02$\;mm, and $\Delta t_0=0.1$\;s (for a domain radius $R_{max}=7$\;mm and a treatment duration of 30\;d) resulted, with a run-time of $\sim0.5$\;s, in doses which were $\sim0.5\%$ away from those obtained with $\epsilon_{tol}=10^{-4}$, $\Delta r=0.01$\;mm and $\Delta t_0=0.1$\;s, with a run-time of $\sim3$\;min (both on a modern laptop computer with an Intel i7 processor and 16 GB RAM memory). The latter, more accurate run, was within $7\cdot 10^{-4}$ of the 0D approximation for the $^{220}$Rn+$^{216}$Po dose. 

\begin{figure}[h]
    \includegraphics[scale=0.17,width=0.5\textwidth]{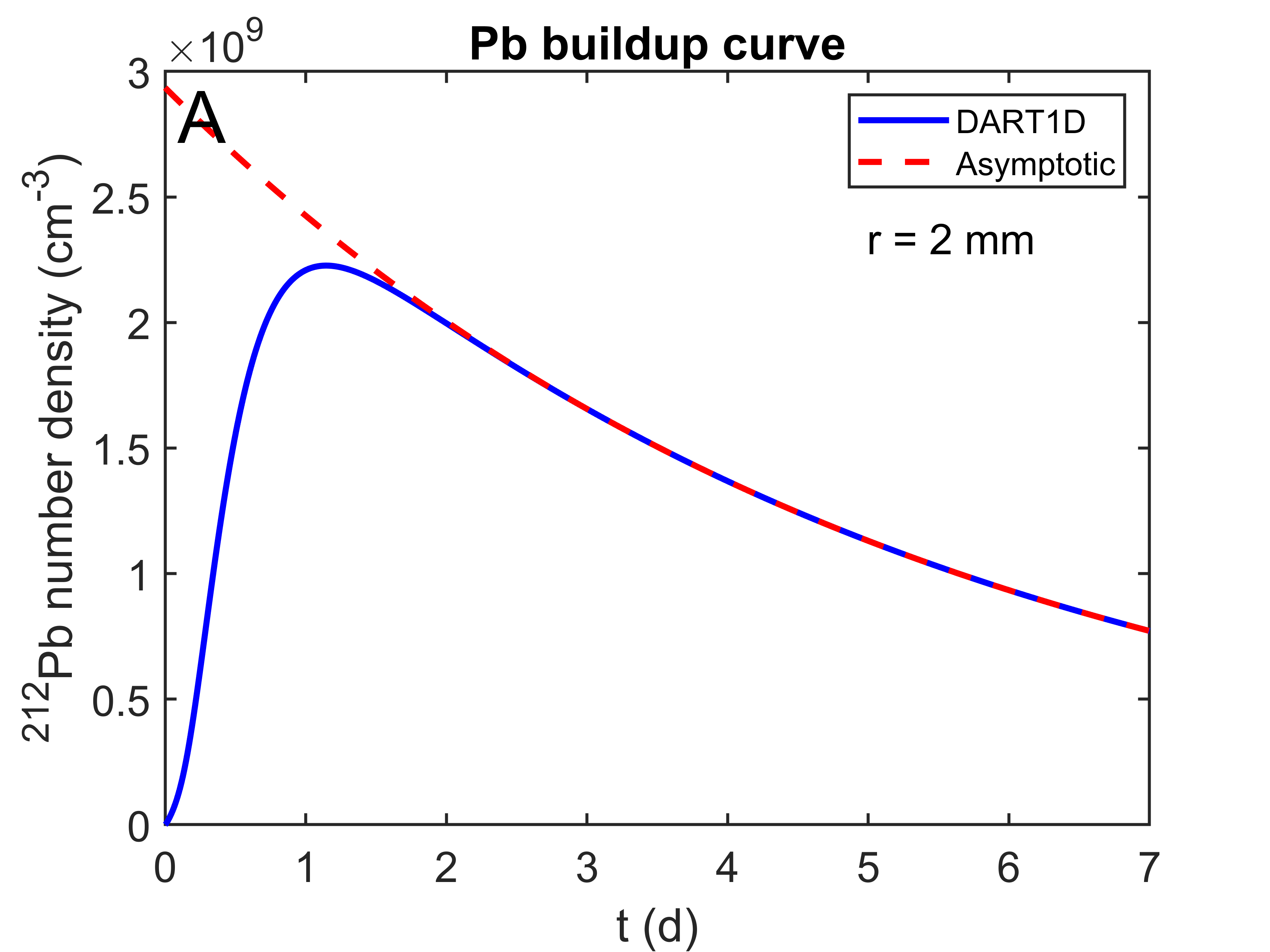}
    \includegraphics[scale=0.2,width=0.5\textwidth]{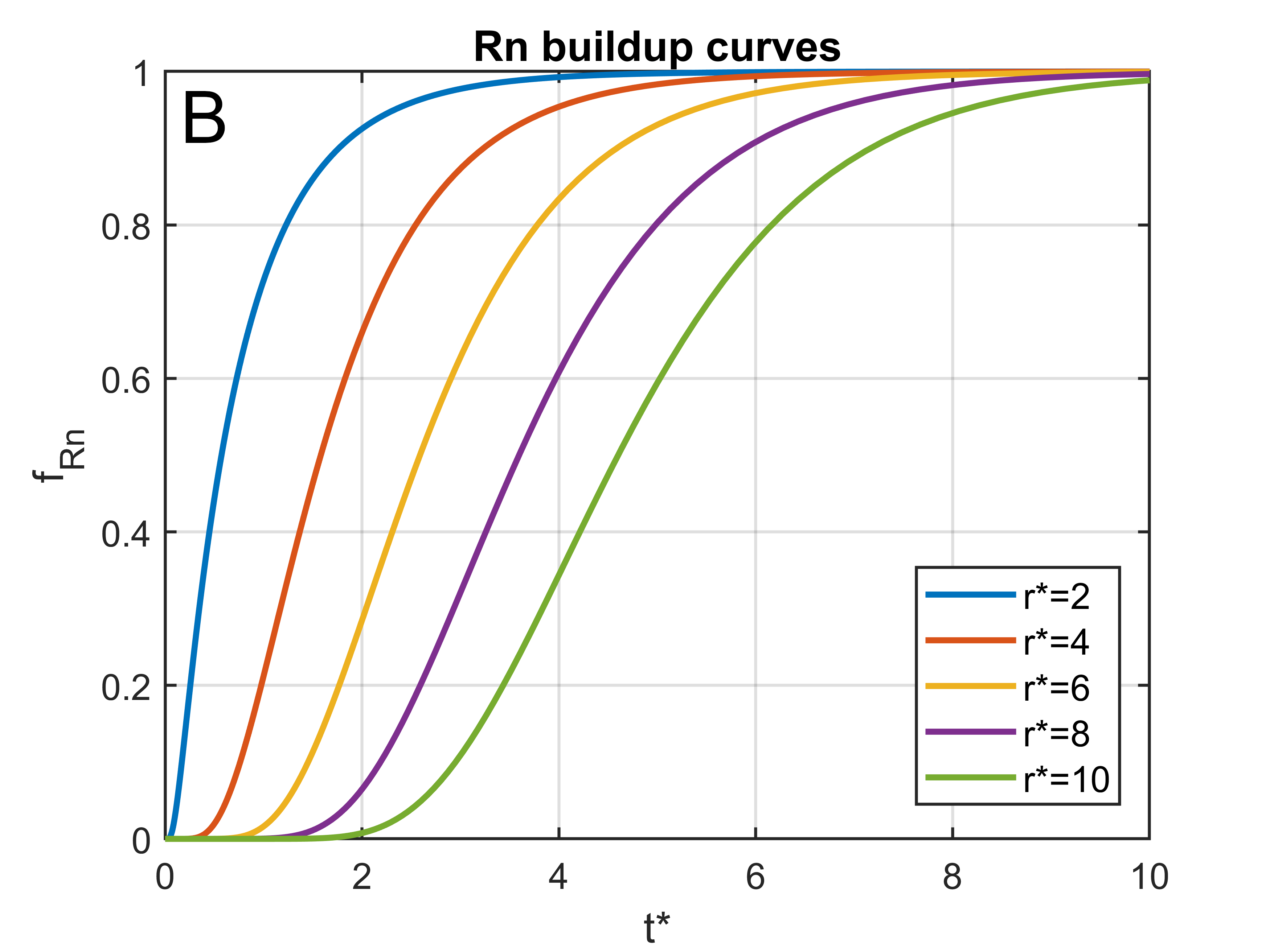}
    \caption{Comparison between the DART1D and asymptotic solutions for the $^{212}$Pb number density 2 mm from the source (A) and the ratio between the DART1D and asymptotic solutions of the $^{220}$Rn number density at various distances from the source axis (B). The distance and time are normalized to $^{220}$Rn diffusion length and mean lifetime: $r^*\equiv r/L_{Rn}$, $t^*\equiv (\lambda_{Rn}-\lambda_{Ra})t$.}
    \label{fig:Fig2_buildupCurves}
\end{figure}

\begin{figure}
    \centering
    \includegraphics[scale=0.15,width=0.49\textwidth]{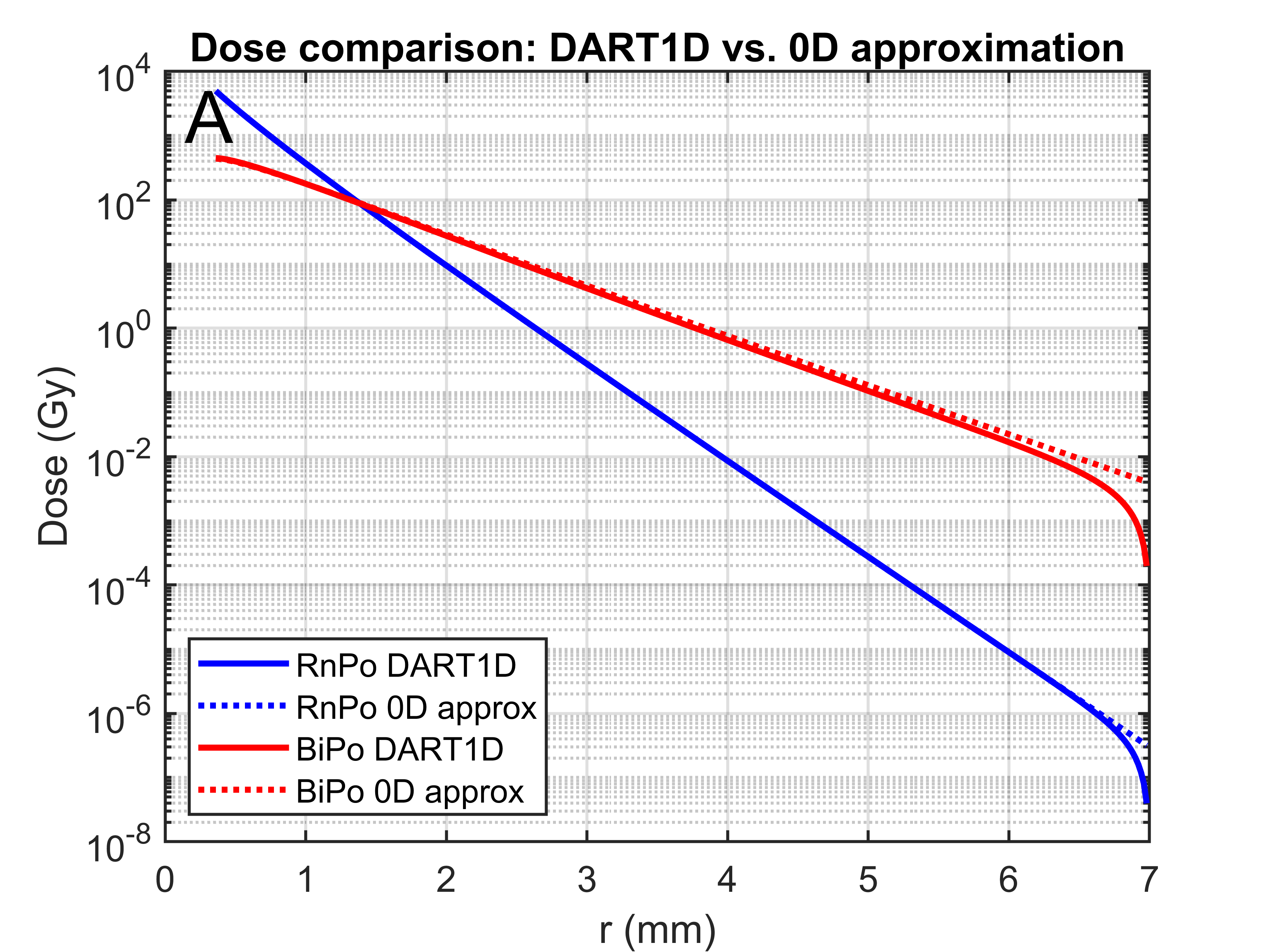}
    \includegraphics[scale=0.15,width=0.49\textwidth]{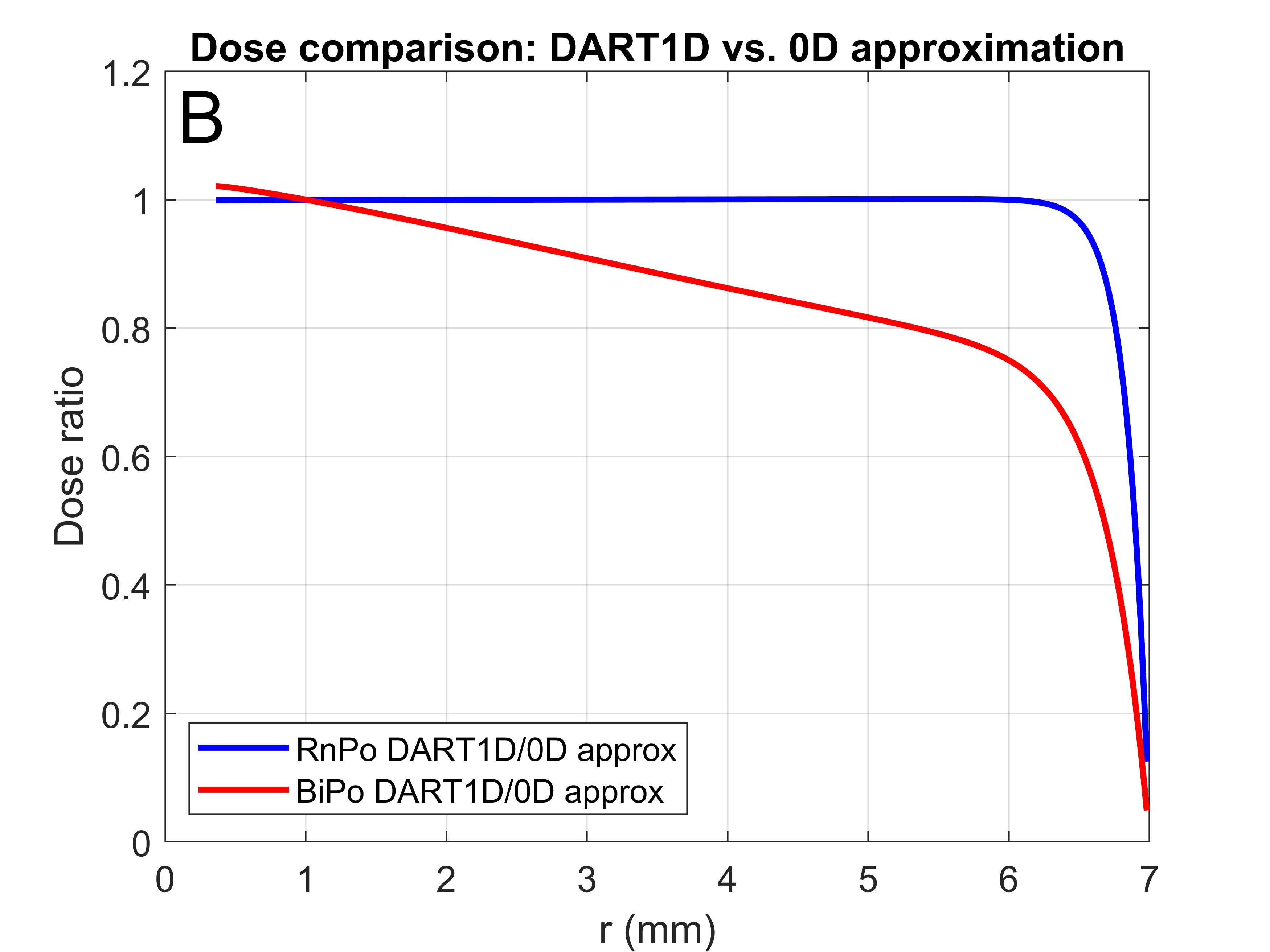}
    \caption{Comparison between the DART1D $^{220}$Rn+$^{216}$Po and $^{212}$Bi/$^{212}$Po alpha doses and the 0D approximation for an infinite cylindrical source. (A) Dose values, (B) DART1D/0D approximation ratios.}
    \label{fig:Fig3_DART1D_vs_0D}
\end{figure}

Figures \ref{fig:Fig2_buildupCurves} and \ref{fig:Fig3_DART1D_vs_0D} examine several aspects of the numerical solution. Figure \ref{fig:Fig2_buildupCurves} shows the DART1D solution in comparison with the asymptotic expressions eq. (\ref{eq:nRn_asy}) and (\ref{eq:nPb_asy}). On the left we show the DART1D time-dependent $^{212}$Pb number density at a distance of 2\;mm from the source, along with the corresponding asymptotic solution. On the right, we show the ratio between the numerical and asymptotic solutions for $^{220}$Rn, $f_{Rn} \equiv n_{Rn}/n_{Rn}^{asy}$, plotted for varying distances from the source. The distances are given in units of the $^{220}$Rn diffusion length, $r^*\equiv r/L_{Rn}$, and the time in units of $1/(\lambda_{Rn}-\lambda_{Ra})$ which is roughly the mean $^{220}$Rn lifetime, $t^*\equiv (\lambda_{Rn}-\lambda_{Ra})t$. The numerical solutions converge to the asymptotic ones with a delay that increases with the distance from the source. For $^{220}$Rn this occurs on the scale of minutes, while for $^{212}$Pb -- over a few days. The adaptive time step allows DART1D to handle both transients efficiently. 

Figure \ref{fig:Fig3_DART1D_vs_0D}A shows the DART1D $^{220}$Rn+$^{216}$Po and $^{212}$Bi/$^{212}$Po alpha dose components calculated for the case $L_{Rn}=0.3$\;mm, $L_{Pb}=0.6$\;mm, $L_{Bi}=0.1L_{Pb}$, $\alpha_{Pb}=\lambda_{Pb}$ (i.e., $P_{leak}(Pb)=0.5$), and $\alpha_{Bi}=0$. The source radius is $R_0=0.35$\;mm, the $^{224}$Ra activity is 3\;$\mu$Ci/cm and the desorption probabilities are $P_{des}(Rn)=0.45$ and $P_{des}^{eff}(Pb)=0.55$. The dose components are given at $t=30$\;d post treatment. The numerical calculation is compared to the 0D approximations, eq. (\ref{eq:Dose(RnPo)_asy_0D}) and (\ref{eq:Dose(BiPo)_asy_0D}). The assumption of zero number density outside the calculation domain results in a departure from the expected solution about two diffusion lengths away from the boundary: $\sim0.5$\;mm for $^{220}$Rn and $\sim1$\;mm for $^{212}$Bi and $^{212}$Po, whose spatial distribution is governed by the $^{212}$Pb diffusion length. This indicates that the radial extent of the calculation domain should be about 10 times larger than the largest diffusion length of the problem. Figure \ref{fig:Fig3_DART1D_vs_0D}B shows the ratio between the DART1D-calculated dose components and the corresponding 0D approximations. Except for the edge effect at $r\rightarrow R_{max}$, the numerical solution for $^{220}$Rn+$^{216}$Po coincides with the 0D approximation to better than $1\cdot 10^{-3}$ for $\epsilon_{tol}=10^{-4}$, $\Delta r=0.01$\;mm and $\Delta t_0=0.1$\;s up to $r\sim5$\;mm. For $^{212}$Bi/$^{212}$Po the 0D approximation underestimates the dose at $r<1$\;mm and overestimates it at larger distances because of the increasing delay in buildup of $^{212}$Pb as a function of $r$. The error is $\sim5-10\%$ at therapeutically relevant distances from the source (around 2-3 mm).

\section{Finite-element two-dimensional time-dependent solution for a finite cylindrical source}

\subsection{Numerical scheme}

We move now to two dimensions to treat a cylindrical source (``seed'') of radius $R_0$ and finite length $l$. The source lies along the $z$ axis with $z=0$ at its mid plane. We solve the DL equations over a cylindrical domain extending from $r=0$ to $r=R_{max}$ and from $z=-Z_{max}$ to $z=+Z_{max}$. Both $R_{max}$ and $Z_{max}-l/2$ should be much larger than the largest diffusion length of the problem. The domain comprises ring elements of equal radial width $\Delta r$ and equal $z$-width $\Delta z$. We choose $\Delta r$ and $\Delta z$ such that $R_0/\Delta r$ and $l/(2\Delta z)$ are integer numbers, with $\Delta r$ and $\Delta z$ much smaller than $L_{Rn}$ and $L_{Pb}$. We enumerate the rings by $i,j$, where $i=1...N_r$ and $j=1...N_z$. Elements with $i=1$ are on-axis, while $j=1$ at the bottom of the cylindrical domain, and $j=N_z$ at the top. Unlike the 1D case, where the source is infinitely long and we only consider points with $r>R_0$, for a finite seed in 2D we must also solve the equations for points above and below the seed, with $r<R_0$ and $|z|>\frac{1}{2} l$. As for the 1D case, the radius and $z$-coordinate of the $i,j$ ring, $r_i,z_j$, are defined at the center of its $rz$ cross section. For the innermost $i=1$ rings $r_1=\frac{1}{2}\Delta r$. Points inside the seed, i.e., in rings with $r_i \leq R_0-\Delta r/2$, and $|z_j| \leq \frac{1}{2} l - \frac{1}{2} \Delta z$, have zero number densities of $^{220}$Rn, $^{212}$Pb and $^{212}$Bi.

\bigskip

Discretization of eq. (\ref{eq:Rn_diffusionEqn_2D}-\ref{eq:Bi_diffusionEqn_2D}) in 2D yields, for interior ring elements in the cylindrical domain (outside of the seed and not touching its wall or bases, and with $i>1$): 
\bigskip

\begin{equation} 
\eqalign{
        \frac{n^{(p+1)}_{x,i,j}-n^{(p)}_{x,i,j}}{\Delta t} = \cr D_{x}\bigg(\frac{n^{(p+1)}_{x,i+1,j}+n^{(p+1)}_{x,i-1,j}-2n^{(p+1)}_{x,i,j}}{\Delta r^2}
        +\frac{1}{r_i} \frac{n^{(p+1)}_{x,i+1,j}-n^{(p+1)}_{x,i-1,j}}{2\Delta r} \cr +\frac{n^{(p+1)}_{x,i,j+1}+n^{(p+1)}_{x,i,j-1}-2n^{(p+1)}_{x,i,j}}{\Delta z^2}\bigg)-(\lambda_{x} + \alpha_{x}) n^{(p+1)}_{x,i,j} + s_{x,i,j}^{(p+1)} \label{eq:diffusionEqn2Ddisc}
        }
\end{equation}
\bigskip

Eq. (\ref{eq:diffusionEqn2Ddisc}) holds also for ring elements on the external surfaces of the domain, with $i=N_r$, $j=1$ or $j=N_z$, as the number densities for rings with $i=N_r+1$, $j=0$ or $j=N_z+1$ are all zero. For ring elements on-axis ($i=1$), above or below the seed, we require $\left(\partial n_x/\partial r\right)_{r=0}=0$. Since the $^{224}$Ra activity is confined to the seed wall, the $z$-component of the current density is set to zero on the seed bases, i.e., $\left(\partial n_x\partial z\right)_{z=\pm l/2}=0$. Defining $i_s$ as the radial index of ring elements touching the seed wall (i.e., $r_{i_s}=R_0+\Delta r/2$), for ring elements with $|z_j|\leq l/2-\Delta z/2$, eq. (\ref{eq:diffusionEqn2Ddisc}) becomes, similarly to the 1D case:

\begin{eqnarray} \label{eq:2D_eq_first_shell}
\fl \qquad \frac{n^{(p+1)}_{x,i_s,j}-n^{(p)}_{x,i_s,j}}{\Delta t}=&\frac{D_x}{\Delta r^2}\left(\frac{1+\Delta r/R_0}{1+\Delta r/2R_0} \right)\left(n^{(p+1)}_{x,i_s+1,j}-n^{(p+1)}_{x,i_s,j}\right) + \nonumber \\
\fl \qquad &\frac{D_x}{\Delta z^2}\left(n^{(p+1)}_{x,i_s,j+1}+n^{(p+1)}_{x,i_s,j-1}-2n^{(p+1)}_{x,i_s,j}\right) -(\lambda_{x}+\alpha_{x})n^{(p+1)}_{x,i_s,j}+s^{(p+1)}_{x,i_s,j} \nonumber \\ 
\end{eqnarray}

The source terms in eq. (\ref{eq:diffusionEqn2Ddisc}) and (\ref{eq:2D_eq_first_shell}) are similar to the 1D case, with the additional requirement that $|z_j|<l/2$:   

\begin{equation}
\fl \qquad s_{Rn,i,j}^{p+1} = \frac{P_{des}(Rn)\left(\Gamma_{Ra}^{src}(0)/l\right) e^{-\lambda_{Ra}t_{p+1}} }{2\pi R_0\Delta r\left(1+\Delta r/2R_0\right) } \delta_{i,i_s} \cdot \left(\frac{1-\mathrm{sign}(|z_j|-l/2)}{2}\right)
\end{equation}

\begin{eqnarray}
\fl \qquad s_{Pb,i,j}^{p+1} = \frac{\left(P_{des}^{eff}(Pb)-P_{des}(Rn)\right)\left(\Gamma_{Ra}^{src}(0)/l\right) e^{-\lambda_{Ra}t_{p+1}} }{2\pi R_0\Delta r\left(1+\Delta r/2R_0\right)}\delta_{i,i_s} \cdot \left(\frac{1-\mathrm{sign}(|z_j|-l/2)}{2}\right) \nonumber \\ \nonumber \\  +\lambda_{Rn}n_{Rn,i,j}^{p+1}
\end{eqnarray}

\begin{equation}
\fl \qquad s_{Bi,i,j}^{p+1} = \lambda_{Pb}n_{Pb,i,j}^{p+1} 
\end{equation}
In order to solve eq. (\ref{eq:diffusionEqn2Ddisc}) in matrix form we use linear indexing. We rearrange the 2D elements $n_{x,i,j}^{(p)}$ and $s_{x,i,j}^{(p)}$ in two column vectors $\mathbf{\widetilde{n}}_x^{(p)}$ and $\mathbf{\widetilde{s}}_x^{(p)}$ in sequential order. 
We define:

\begin{equation}
    k(i,j)=(j-1)N_r + i
\end{equation}

\begin{equation}
    \widetilde{n}_{x,k}^{(p)} = n_{x,i,j}^{(p)}
\end{equation}

\begin{equation}
    \widetilde{s}_{x,k}^{(p)} = s_{x,i,j}^{(p)}
\end{equation}
with $k=1...N_r Nz$. Noting that $n_{x,i\pm1,j}^{(p)}=\widetilde{n}_{x,k\pm1}^{(p)}$ and $n_{x,i,j\pm1}^{(p)}=\widetilde{n}_{x,k\pm N_r}^{(p)}$, eq. (\ref{eq:diffusionEqn2Ddisc}) can be rearranged as: 

\begin{eqnarray} \label{eq:diffusionEqn2Ddisc_reshaped}
\fl \quad \widetilde{n}^{(p)}_{x,k}+\widetilde{s}_{x,k}^{(p+1)}\Delta t = \nonumber \\ 
\fl \quad M_{k,k-N_r}^{(x)}\widetilde{n}^{(p+1)}_{x,k-N_r} + M_{k,k-1}^{(x)}\widetilde{n}^{(p+1)}_{x,k-1} + M_{k,k}^{(x)}\widetilde{n}^{(p+1)}_{x,k} + M_{k,k+1}^{(x)}\widetilde{n}^{(p+1)}_{x,k+1} + M_{k,k+N_r}^{(x)}\widetilde{n}^{(p+1)}_{x,k+N_r}
\end{eqnarray}

As for the 1D case, the matrix elements appearing in eq. (\ref{eq:diffusionEqn2Ddisc_reshaped}) depend on the values of $i,j$ (and therefore $k$), in a manner that satisfies the boundary conditions. For compactness, we define the following intermediate quantities:

\begin{eqnarray}
    K_z^{(x)} &= \frac{D_x\Delta t}{\Delta z^2} \\
    K_r^{(x)} &= \frac{D_x\Delta t}{\Delta r^2} \\
    K_r^{(x+)} &= \frac{D_x\Delta t}{\Delta r^2}\left(1+\frac{\Delta r}{2r_i}\right) \\
    K_r^{(x-)} &= \frac{D_x\Delta t}{\Delta r^2}\left(1-\frac{\Delta r}{2r_i}\right) \\
    S_1^{(x)} &= 1 + 2K_r^{(x)} + K_z^{(x)} + (\lambda_x+\alpha_x)\Delta t \\
    S_2^{(x)} &= 1 + 2K_r^{(x)} + 2K_z^{(x)} + (\lambda_x+\alpha_x)\Delta t \\
    S_+^{(x)} &= 1 + K_r^{(x+)} + 2K_z^{(x)} + (\lambda_x+\alpha_x)\Delta t
\end{eqnarray}
Table \ref{tab:M2D_cases} lists the expressions for the matrix elements $M_{k,l}^{(x)}$ for all possible cases for $r_i$ and $z_j$. With these, we can write eq. (\ref{eq:diffusionEqn2Ddisc_reshaped}) in matrix form (with $K\equiv N_rN_z$):

\begin{table}
\caption{\label{tab:M2D_cases}Matrix elements in 2D.}
\begin{tabular}{l | c c c c c}
\br
Case                                                                   &$M_{k,k-N_r}^{(x)}$ &$M_{k,k-1}^{(x)}$ &$M_{k,k}^{(x)}$ &$M_{k,k+1}^{(x)}$ &$M_{k,k+N_r}^{(x)}$\\
\br
$R_0+\frac{\Delta r}{2} < r_i < R_{max}-\frac{\Delta r}{2}$            &                    &                  &                &                  &                   \\
\& $|z_j|<Z_{max}-\frac{\Delta z}{2}$                                  &$-K_z^{(x)}$        &$-K_r^{(x-)}$     &$S_2^{(x)}$     &$-K_r^{(x+)}$     &$-K_z^{(x)}$       \\
\mr
$\frac{\Delta r}{2} < r_i \leq R_0+\frac{\Delta r}{2}$                 &                    &                  &                &                  &                   \\
\& $\frac{l}{2}+\frac{\Delta z}{2}<|z_j|<Z_{max}-\frac{\Delta z}{2}$   &$-K_z^{(x)}$        &$-K_r^{(x-)}$     &$S_2^{(x)}$     &$-K_r^{(x+)}$     &$-K_z^{(x)}$       \\
\mr
$r_i=R_0+\frac{\Delta r}{2}$ \& $|z_j|=\frac{l}{2}+\frac{\Delta z}{2}$ &$-K_z^{(x)}$        &$-K_r^{(x-)}$     &$S_2^{(x)}$     &$-K_r^{(x+)}$     &$-K_z^{(x)}$       \\
\mr
$r_i=R_0+\frac{\Delta r}{2}$ \& $|z_j|<\frac{l}{2}$                    &$-K_z^{(x)}$        &0                 &$S_+^{(x)}$     &$-K_r^{(x+)}$     &$-K_z^{(x)}$       \\
\mr
$\frac{\Delta r}{2}<r_i<R_0$ \& $z_j=\frac{l}{2}+\frac{\Delta z}{2}$   &0                   &$-K_r^{(x-)}$     &$S_1^{(x)}$     &$-K_r^{(x+)}$     &$-K_z^{(x)}$       \\
\mr
$\frac{\Delta r}{2}<r_i<R_0$ \& $z_j=-\frac{l}{2}-\frac{\Delta z}{2}$  &$-K_z^{(x)}$        &$-K_r^{(x-)}$     &$S_1^{(x)}$     &$-K_r^{(x+)}$     &0                  \\
\mr
$r_i=\frac{\Delta r}{2}$                                               &                    &                  &                &                  &                   \\
\& $\frac{l}{2}+\frac{\Delta z}{2}<|z_j|<Z_{max}-\frac{\Delta z}{2}$   &$-K_z^{(x)}$        &0                 &$S_2^{(x)}$     &$-K_r^{(x+)}$     &$-K_z^{(x)}$       \\
\mr
$r_i=R_{max}-\frac{\Delta r}{2}$                                       &                    &                  &                &                  &                   \\
\& $|z_j|<Z_{max}-\frac{\Delta z}{2}$                                  &$-K_z^{(x)}$        &$-K_r^{(x-)}$     &$S_2^{(x)}$     &0                 &$-K_z^{(x)}$       \\
\mr
$\frac{\Delta r}{2}<r_i<R_{max}-\frac{\Delta r}{2}$                    &                    &                  &                &                  &                   \\
\& $z_j=Z_{max}-\frac{\Delta z}{2}$                                    &$-K_z^{(x)}$        &$-K_r^{(x-)}$     &$S_2^{(x)}$     &$-K_r^{(x+)}$     &0                  \\
\mr
$\frac{\Delta r}{2}<r_i<R_{max}-\frac{\Delta r}{2}$                    &                    &                  &                &                  &                   \\
\& $z_j=-Z_{max}+\frac{\Delta z}{2}$                                   &0                   &$-K_r^{(x-)}$     &$S_2^{(x)}$     &$-K_r^{(x+)}$     &$-K_z^{(x)}$       \\
\mr
$r_i=\frac{\Delta r}{2} \;\&\; z_j=Z_{max}-\frac{\Delta z}{2}$         &$-K_z^{(x)}$        &0                 &$S_2^{(x)}$     &$-K_r^{(x+)}$     &0                  \\
\mr
$r_i=\frac{\Delta r}{2} \;\&\; z_j=-Z_{max}+\frac{\Delta z}{2}$        &0                   &0                 &$S_2^{(x)}$     &$-K_r^{(x+)}$     &$-K_z^{(x)}$       \\
\mr
$r_i=\frac{\Delta r}{2} \;\&\; z_j=\frac{l}{2}+\frac{\Delta z}{2}$     &0                   &0                 &$S_2^{(x)}$     &$-K_r^{(x+)}$     &$-K_z^{(x)}$       \\
\mr
$r_i=\frac{\Delta r}{2} \;\&\; z_j=-\frac{l}{2}-\frac{\Delta z}{2}$    &$-K_z^{(x)}$        &0                 &$S_2^{(x)}$     &$-K_r^{(x+)}$     &0                  \\
\mr
$r_i=R_{max}-\frac{\Delta r}{2}$ \& $z_j=Z_{max}-\frac{\Delta z}{2}$   &$-K_z^{(x)}$        &$-K_r^{(x-)}$     &$S_2^{(x)}$     &0                 &0                  \\
\mr
$r_i=R_{max}-\frac{\Delta r}{2}$ \& $z_j=-Z_{max}+\frac{\Delta z}{2}$  &0                   &$-K_r^{(x-)}$     &$S_2^{(x)}$     &0                 &$-K_z^{(x)}$       \\
\mr
$r_i<R_{0}$ \& $|z_j|<\frac{l}{2}$                                     &0                   &0                 &1               &0                 &0                  \\
\br
\end{tabular}\\
\end{table}

\begin{eqnarray} \label{A_matrix2D}
\fl
\resizebox{1\hsize}{!}{$
\left( \matrix{
    \medskip \widetilde{n}^{(p)}_{x,1} \cr 
    \medskip \widetilde{n}^{(p)}_{x,2} \cr 
    \widetilde{n}^{(p)}_{x,3} \cr 
    \vdots \cr
    \cr
    \medskip \widetilde{n}^{(p)}_{x,N_r+1} \cr
    \widetilde{n}^{(p)}_{x,N_r+2} \cr
    \cr
    \vdots\cr
    \cr
    \medskip \widetilde{n}^{(p)}_{x,K-1} \cr
    \widetilde{n}^{(p)}_{x,K} 
} \right)
+
\left( \matrix{
    \medskip \widetilde{s}^{(p+1)}_{x,1} \cr
    \medskip \widetilde{s}^{(p+1)}_{x,2} \cr
    \widetilde{s}^{(p+1)}_{x,3} \cr
    \vdots \cr
    \cr
    \medskip \widetilde{s}^{(p+1)}_{x,N_r+1} \cr
    \widetilde{s}^{(p+1)}_{x,N_r+2} \cr
    \cr
    \vdots\cr
    \cr
    \medskip \widetilde{s}^{(p+1)}_{x,K-1} \cr
    \widetilde{s}^{(p+1)}_{x,K} 
} \right) \Delta t
= 
\left( \matrix{
\medskip M_{1,1}^{(x)}     &M_{1,2}^{(x)} &0             &\cdots        &M_{1,1+N_r}^{(x)} &\cdots            &                  &       &0\cr
\medskip M_{2,1}^{(x)}     &M_{2,2}^{(x)} &M_{2,3}^{(x)} &0             &\cdots            &M_{2,2+N_r}^{(x)} &\cdots            &       &0\cr
         0                 &M_{3,2}^{(x)} &M_{3,3}^{(x)} &M_{3,4}^{(x)} &0                 &\cdots            &M_{3,3+N_r}^{(x)} &\cdots &0\cr
        \vdots             &              &\ddots        &\ddots        &\ddots            &                  &\ddots            &       &\vdots\cr
        \cr
\medskip M_{N_r+1,1}^{(x)} &0                 &\cdots  &M_{N_r+1,N_r}^{(x)} &M_{N_r+1,N_r+1}^{(x)} &M_{N_r+1,N_r+2}^{(x)} &0                     &\cdots  &0\cr
         0                 &M_{N_r+2,2}^{(x)} &0       &\cdot               &M_{N_r+2,N_r+1}^{(x)} &M_{N_r+2,N_r+2}^{(x)} &M_{N_r+2,N_r+3}^{(x)} &\cdots  &0\cr
         \cr
         \vdots            &                  &\ddots  &                    &                      &\ddots                &\ddots                &\ddots  &\vdots\cr
         &                 &                  &        &                    &                      &                      &                      &        &\cr
\medskip 0                 &\cdots &0      &M_{K-1,K-1-Nr}^{(x)} &0                &\cdots    &M_{K-1,K-2}^{(x)} &M_{K-1,K-1}^{(x)}  &M_{K-1,K}^{(x)} \cr
         0                 &       &\cdots &0                    &M_{K,K-Nr}^{(x)} &0         &\cdots            &M_{K,K-1}^{(x)} &M_{K,K}^{(x)}              
} \right) 
\left( \matrix{
    \medskip \widetilde{n}^{(p+1)}_{x,1} \cr
    \medskip \widetilde{n}^{(p+1)}_{x,2} \cr
    \widetilde{n}^{(p+1)}_{x,3} \cr
    \vdots \cr
    \cr
    \medskip \widetilde{n}^{(p+1)}_{x,N_r+1} \cr
    \widetilde{n}^{(p+1)}_{x,N_r+2} \cr
    \cr
    \vdots\cr
    \cr
    \medskip \widetilde{n}^{(p+1)}_{x,K-1} \cr
    \widetilde{n}^{(p+1)}_{x,K} 
} \right)$} \nonumber \\ \nonumber \\
\end{eqnarray}
or, equivalently: $\mathbf{\widetilde{n}}^{(p)}_x+\mathbf{\widetilde{s}}^{(p+1)}_x\Delta t=\mathbf{M}_x \mathbf{\widetilde{n}}^{(p+1)}_x$. As for the 1D case, we multiply on the left by the inverse of $\mathbf{M}_x$, getting $\mathbf{\widetilde{n}}^{(p+1)}_x=\mathbf{M}_x^{-1}(\mathbf{\widetilde{n}}^{(p)}_x+\mathbf{\widetilde{s}}^{(p+1)}_x\Delta t)$. We now run over all possible values of $i,j$ and update $n_{x,i,j}^{(p+1)}=\widetilde{n}_{x,k}^{(p+1)}$. Once the new number densities are known in all ring elements, we update the alpha dose components:

\begin{equation}
    \fl \qquad \qquad Dose_{\alpha}^{(p+1)}(RnPo;\,i,j)=Dose_{\alpha}^{(p)}(RnPo;\,i,j)+\frac{E_{\alpha}(RnPo)}{\rho}\lambda_{Rn} n_{Rn,i,j}^{(p+1)}\Delta t    
\end{equation}

\begin{equation}
    \fl \qquad \qquad Dose_{\alpha}^{(p+1)}(BiPo;\,i,j)=Dose_{\alpha}^{(p)}(BiPo;\,i,j)+\frac{E_{\alpha}(BiPo)}{\rho}\lambda_{Bi} n_{Bi,i,j}^{(p+1)}\Delta t    
\end{equation}

As for the 1D case, the time step can be modified in many ways. Here we chose to update it according to the relative change in the overall activity (sum over all isotopes in all ring elements).

\subsection{DART2D: tests and comparisons}

The 2D numerical scheme described above was implemented in MATLAB in a code named ``DART2D''. The code takes roughly 0.5\;h to run on a modern laptop (Intel i7 processor with 16 GB RAM) for $\Delta r = 0.005$\;mm, $\Delta z = 0.05$\;mm, $\epsilon_{tol}=0.01$, $\Delta t_0=0.1$\;s, $R_{max}=7$\;mm, $Z_{max}=10$\;mm and a treatment time of 40\;d. The most demanding process is the calculation of $\mathbf{M}^{-1}$. Since $\mathbf{M}$ is a sparse diagonal matrix, we used MATLAB's \texttt{spdiags()} function, which reduces memory requirements by saving only the diagonal non-zero elements of $\mathbf{M}$, and allows the code to run more efficiently. 

\begin{figure}
    \centering
    \includegraphics[width=0.5\textwidth]{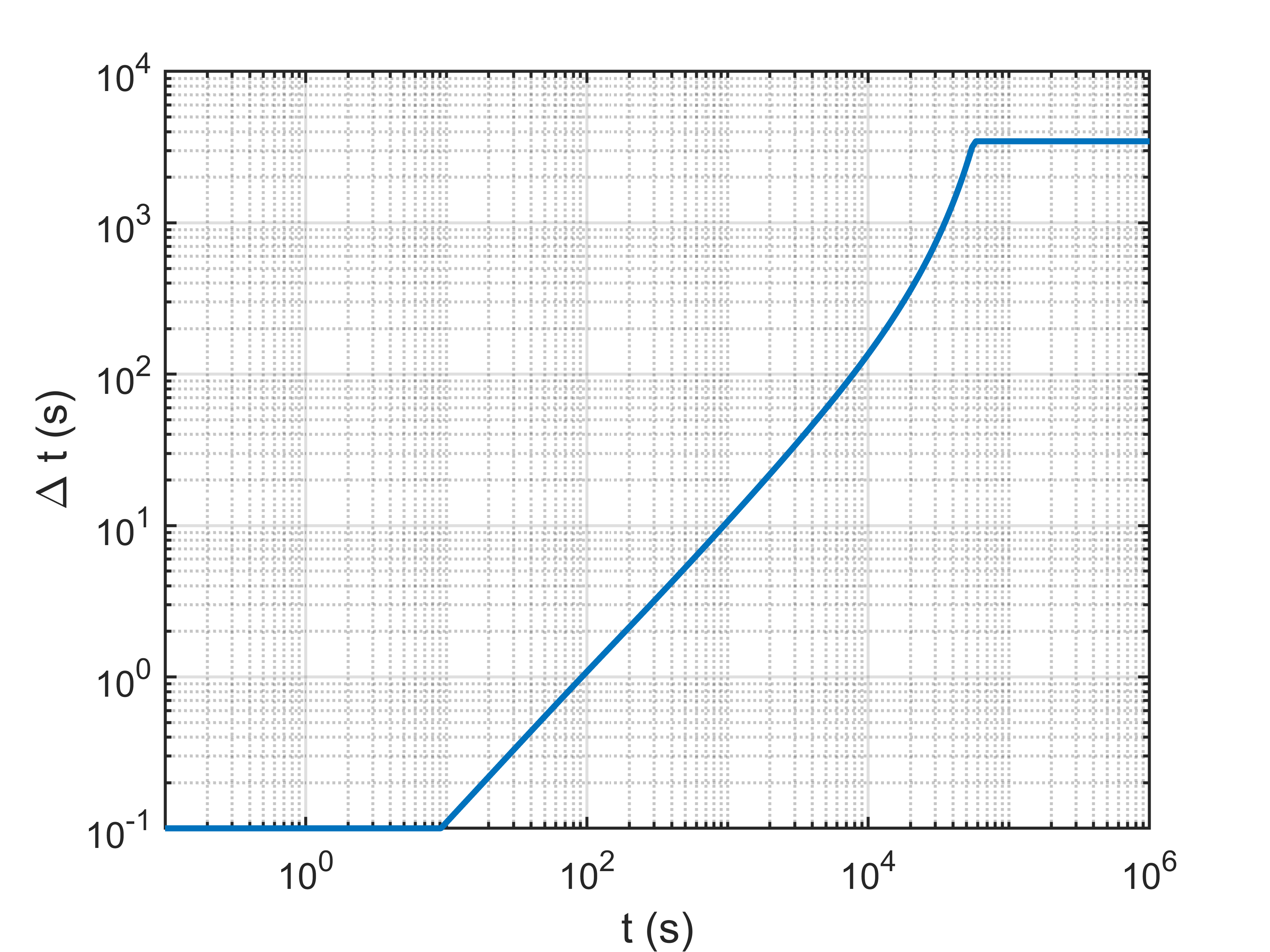}
    \caption{DART2D: variation of the time step vs. time.}
    \label{fig:fig4_Delta_t_vs_t}
\end{figure}

Figure \ref{fig:fig4_Delta_t_vs_t} shows the dependence of the adaptive time step on time, up to $\sim11$\ days. The initial time step is 0.1\;s, capturing the $^{220}$Rn buildup with high accuracy (this is also the minimal allowed value for $\Delta t$). It then gradually increases, following the buildup of $^{212}$Pb, eventually reaching its maximal allowed value (here $\sim1$\;h) in the asymptotic phase driven by $^{224}$Ra decay rate. 

\begin{figure}[h]
    \includegraphics[scale=0.2,width=0.5\textwidth]{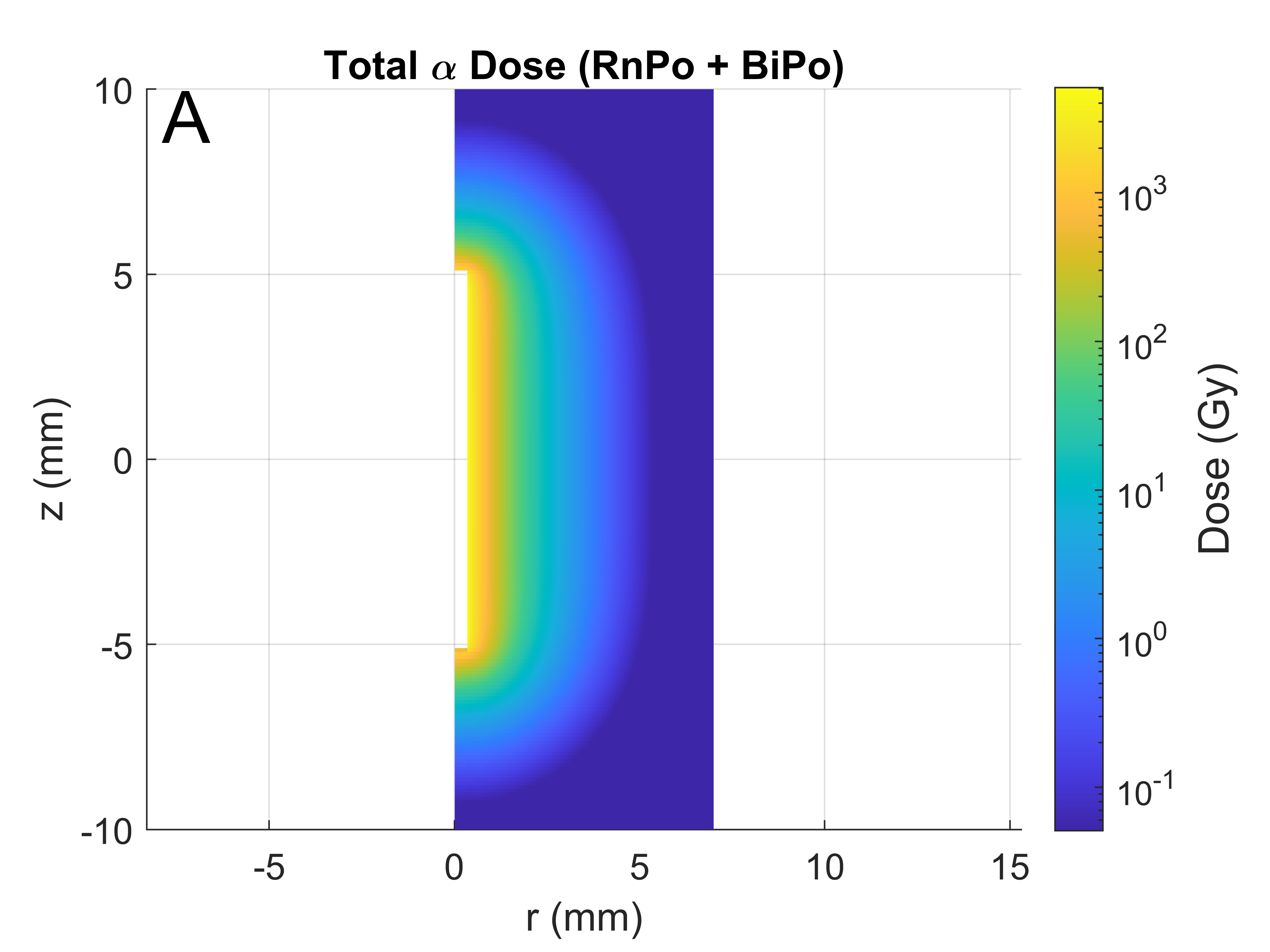}
    \includegraphics[scale=0.2,width=0.5\textwidth]{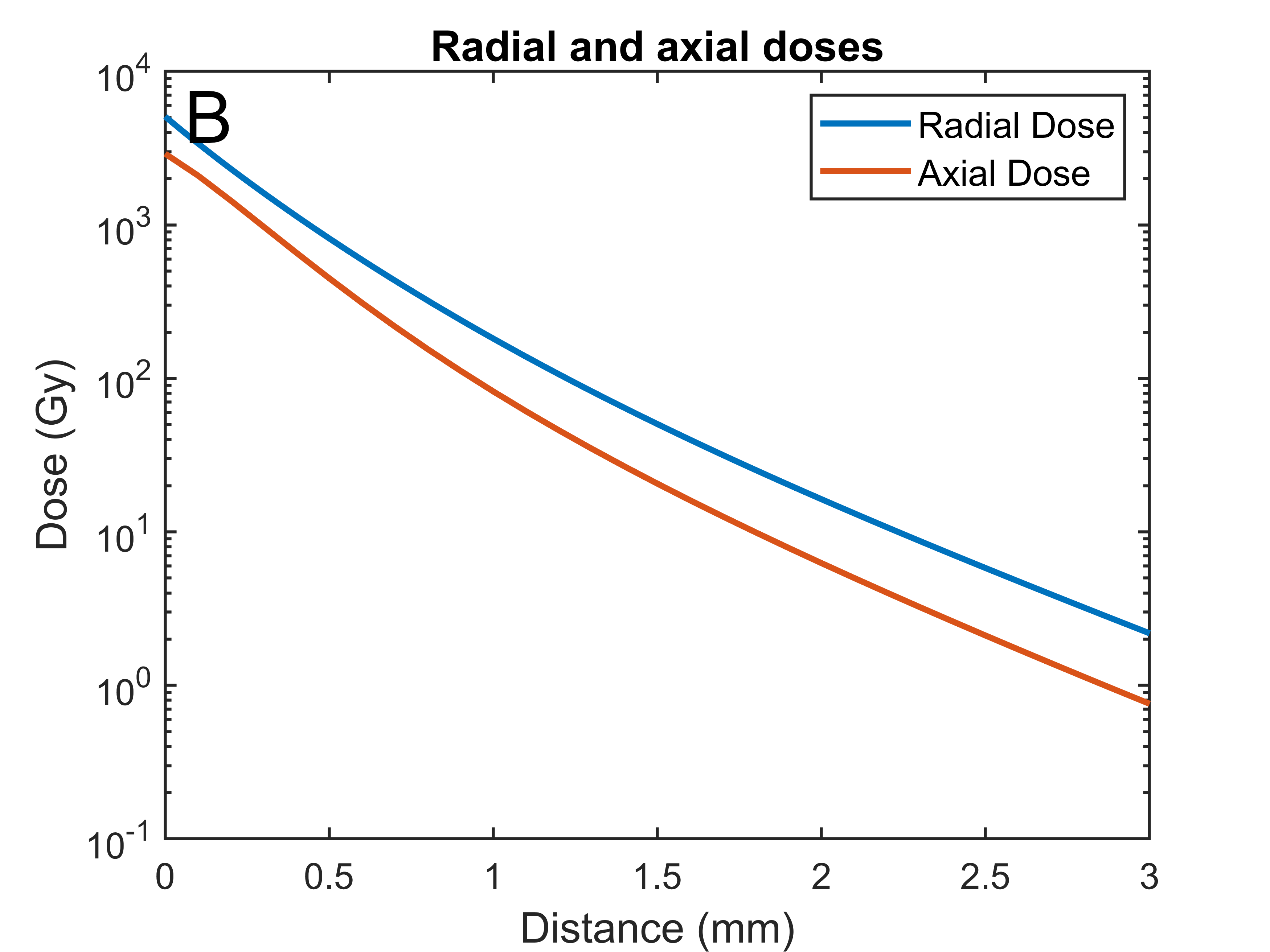}
    \caption{(A) Total alpha dose accumulated over 40 days of treatment by a DaRT seed with initial activity of $3\;\mu$Ci $^{224}$Ra. The seed radius and length are 0.35\;mm and 10\;mm, respectively. The other model parameters are given in the text. (B) Total alpha dose as a function of the distance from the seed edge along $r$ in the mid plane and along $z$ on the seed axis.}
    \label{fig:fig5_2DTotalDose_and_radial_axial_doses}
\end{figure}

The total alpha dose (sum of the $^{220}$Rn+$^{216}$Po and $^{212}$Bi/$^{212}$Po contributions) accumulated over 40 days of DaRT treatment by a seed of finite dimensions is displayed in the $rz$ plane in figure \ref{fig:fig5_2DTotalDose_and_radial_axial_doses}A. The seed dimensions are $R_0=0.35$\;mm and $l=10$\;mm. The initial $^{224}$Ra activity of the seed is 3\;$\mu$Ci, with $P_{des}(Rn) = 0.45$ and $P_{des}^{eff}(Pb) = 0.55$. The other model parameters are: $L_{Pb} = 0.6$\;mm, $L_{Rn} = 0.3$\;mm, $L_{Bi} = 0.06$\;mm, $P_{leak}(Pb) = 0.5$, $\alpha_{Bi}=0$. Note that the radial dose profile is nearly unchanged up to $\sim1.5$\;mm from the seed end. Figure \ref{fig:fig5_2DTotalDose_and_radial_axial_doses}B shows the dose profiles along $r$ in the seed mid plane and along $z$ parallel to the seed axis, both set such that `0' is the seed edge. The dose along the seed axis is smaller by $\sim30\%$ near the seed edge, with the difference increasing to a factor of $\sim3$ at 3 mm, compared to that in the mid plane - an important point to consider in treatment planning. Although a similar effect is observed when approximating the seed to a finite line source comprised of point-like segments, as done in \cite{Arazi2020}, this approach leads to significant errors in the dose because it does not consider the finite diameter of the seed, which ``pushes'' the radial dose to larger values. 

\begin{figure}[h]
    \includegraphics[scale=0.2,width=0.5\textwidth]{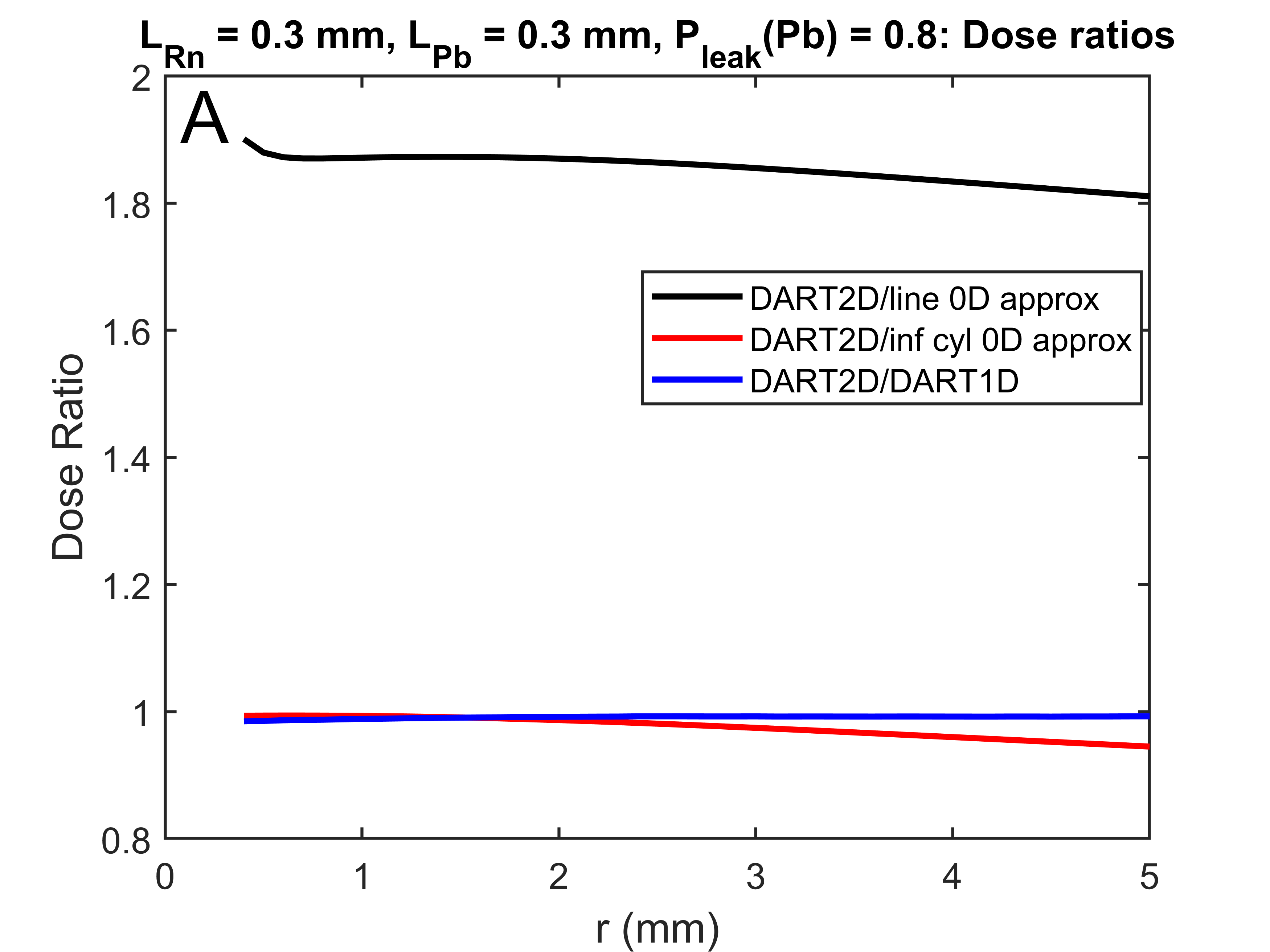}
    \includegraphics[scale=0.2,width=0.5\textwidth]{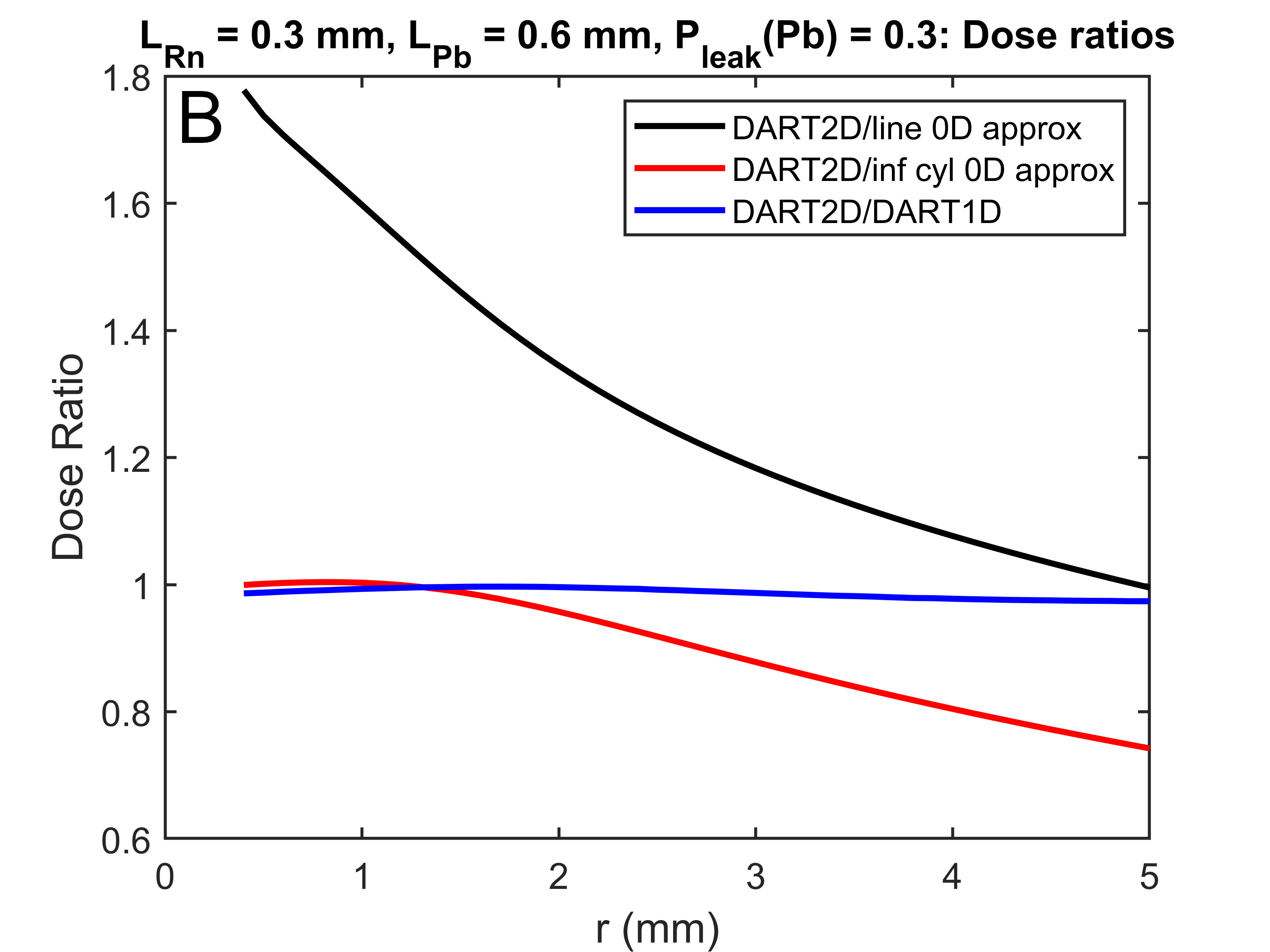}
    \caption{Ratios between the total alpha dose in the seed mid plane calculated by DART2D and those calculated using the 0D line source approximation, the 0D infinite cylinder approximation and the full 1D calculation for an infinite cylindrical source using DART1D. (A) Low-diffusion, high-leakage case; (B) High-diffusion, low leakage.}
    \label{fig:comparisons to 1d and 0d}
\end{figure}

We now compare the results of the full 2D calculation with those obtained using either the 0D analytical approximations, or the full 1D calculation. Figure \ref{fig:comparisons to 1d and 0d} shows these comparisons of the dose profile calculated in the seed mid planes. On the left we display the comparison for a low-diffusion high-leakage case, with $L_{Rn}=0.3$\;mm, $L_{Pb}=0.3$\;mm, and $P_{leak}(Pb)=0.8$, and on the right -- for a high-diffusion low-leakage case, with $L_{Rn}=0.3$\;mm, $L_{Pb}=0.6$\;mm, and $P_{leak}(Pb)=0.3$. In both cases $L_{Bi}=0.1L_{Pb}$ and $\alpha_{Bi}=0$. The curves show the ratios between the full 2D calculation with DART2D and those obtained by: (1) approximating the seed to a finite line comprised of point-like segments and using the 0D approximation, as was done in \cite{Arazi2020}; (2) using the 0D approximation for an infinite cylindrical source, eq.(\ref{eq:Dose(RnPo)_asy_0D}) and (\ref{eq:Dose(BiPo)_asy_0D}), and (3) using the full DART1D calculation. Approximating the seed to a finite line source leads to an underestimation of the dose by up to $\sim 80\%$ for both the low- and high-diffusion scenarios. Using the closed-form 0D approximation for a cylindrical source of radius $R_0$ overestimates the dose at 2-3\;mm by $\sim1-2\%$ for the low-diffusion/high-leakage case and $\sim5-10\%$ for the high-diffusion/low-leakage scenario. In contrast, the full numerical solution (DART1D) for a cylindrical source provides accurate results (on the scale of $0.3\%$) when compared to the 2D calculation.

\begin{figure}[h]
    \includegraphics[scale=0.2,width=0.5\textwidth]{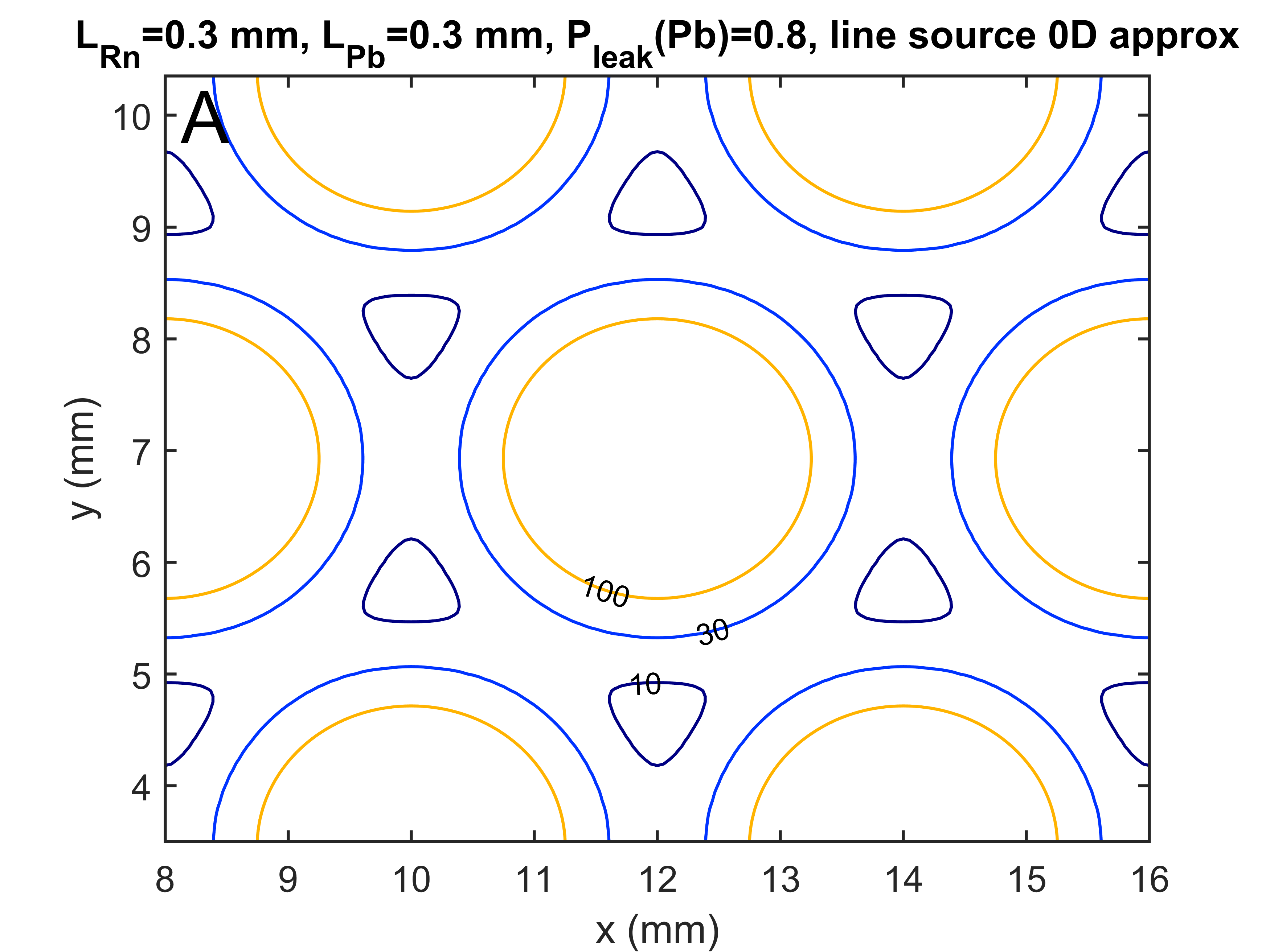}
    \includegraphics[scale=0.2,width=0.5\textwidth]{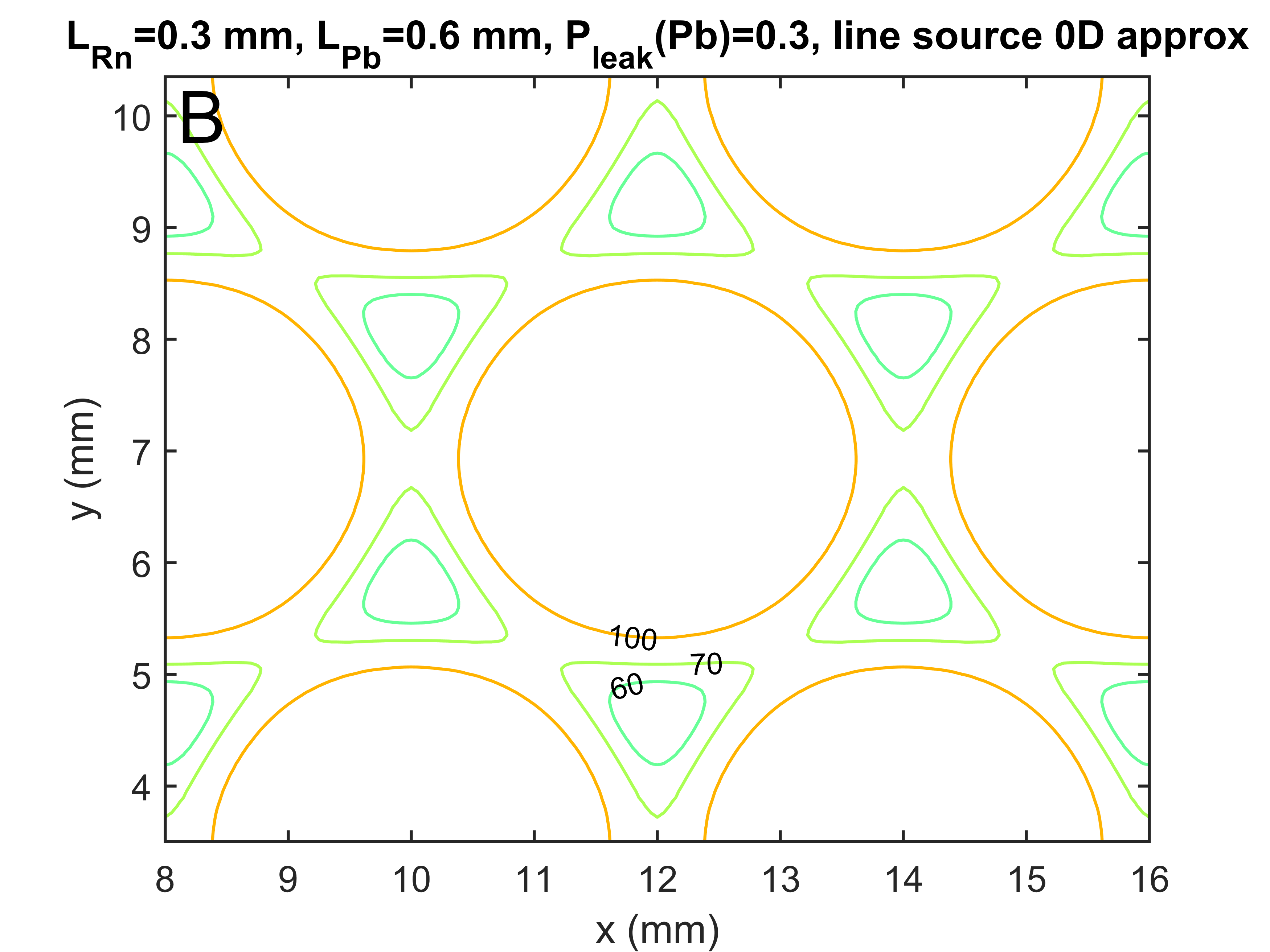}    
    \includegraphics[scale=0.2,width=0.5\textwidth]{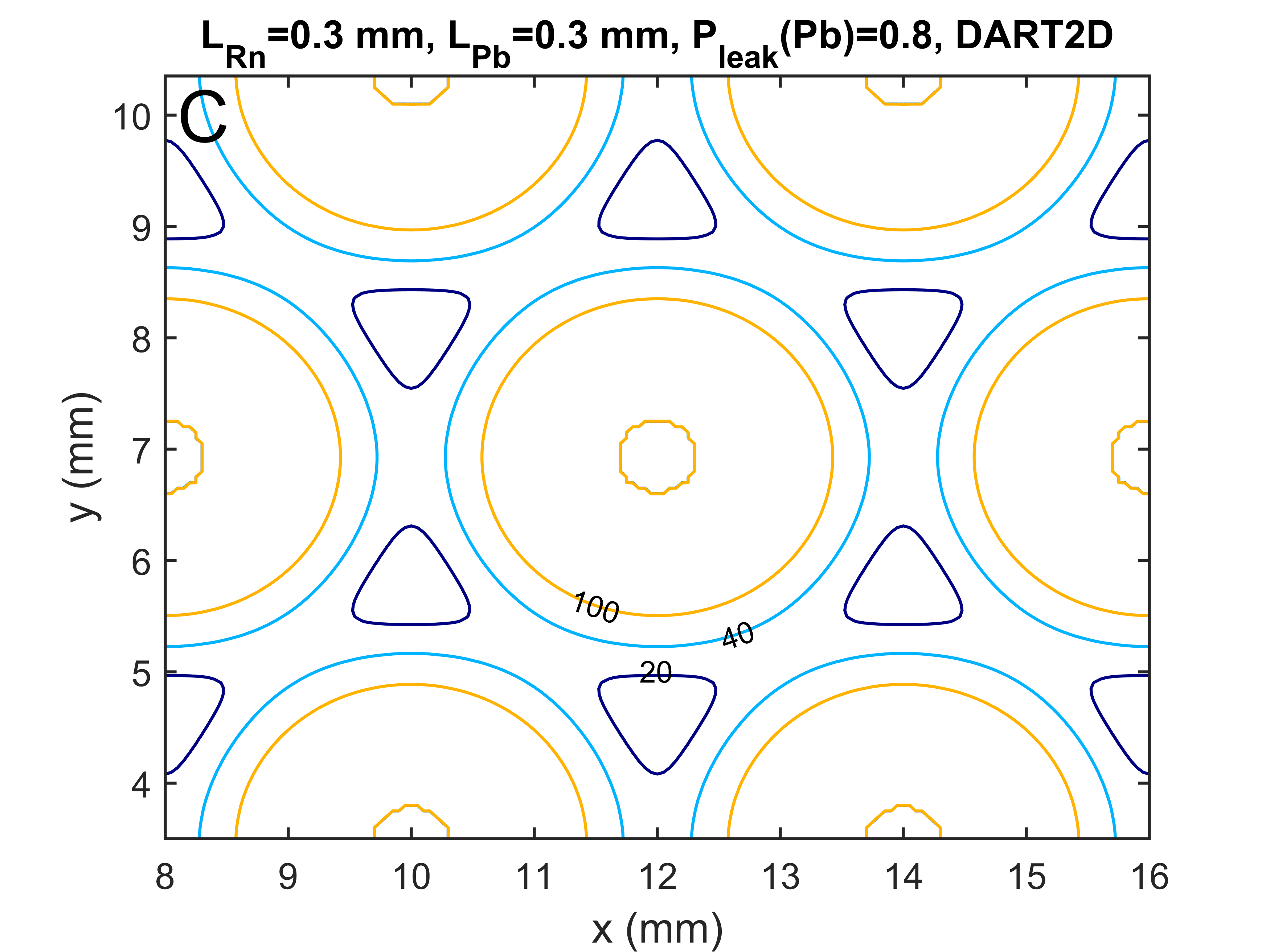}    
    \includegraphics[scale=0.2,width=0.5\textwidth]{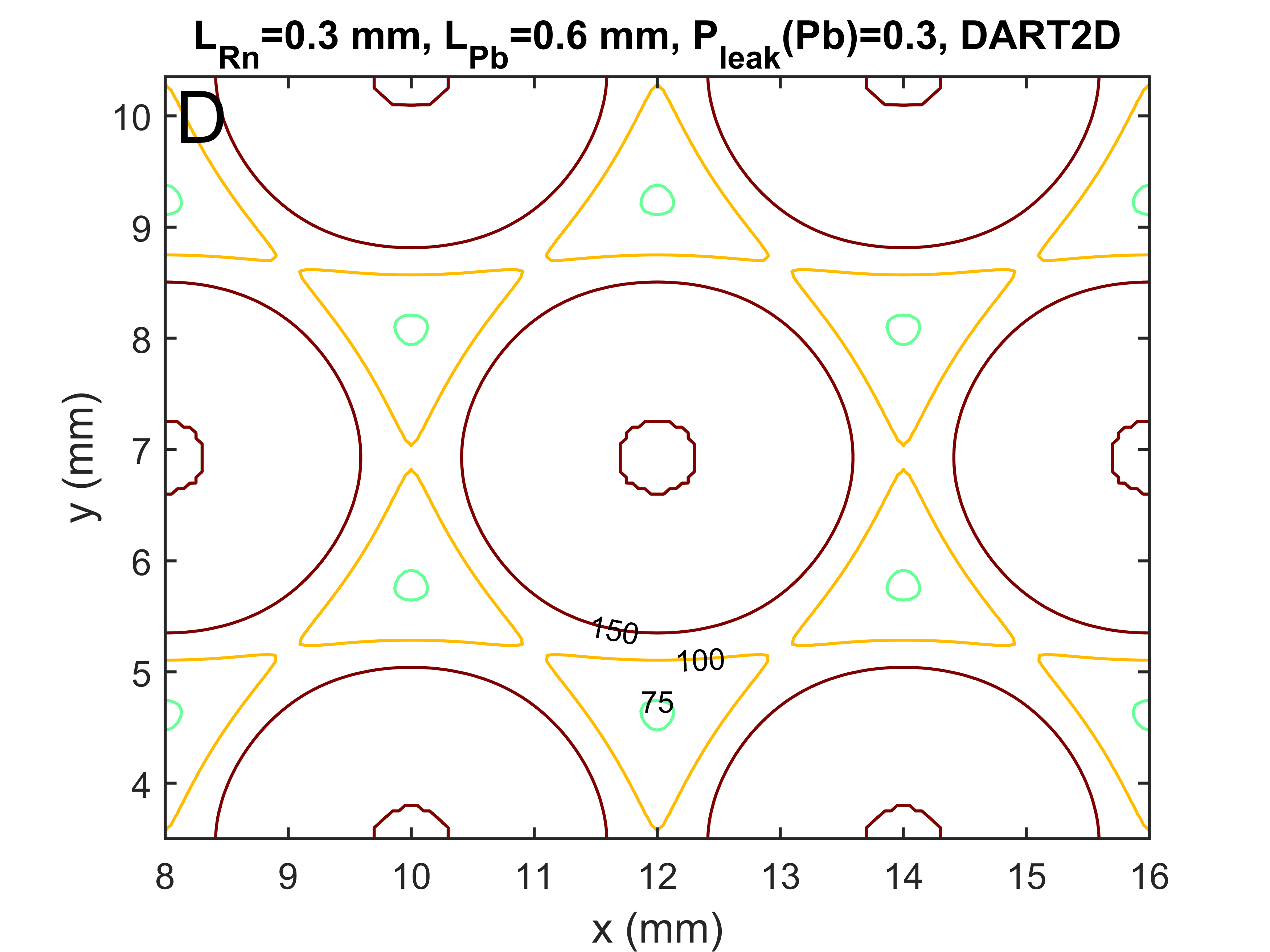}
    \caption{Lattice total alpha dose map comparisons. (A) $L_{Rn} = 0.3$ mm, $L_{Pb} = 0.3$ mm, $P_{leak}(Pb)=0.8$, line source 0D approximation. (B) $L_{Rn} = 0.3$ mm, $L_{Pb} = 0.6$ mm, $P_{leak}(Pb)=0.3$ line source 0D approximation. (C) Same parameters as A, full 2D calculation with DART2D. (D) same parameters as B, full 2D calculation with DART2D. In all figures the grid spacing is 4 mm, seed $^{224}$Ra activity 3\;$\mu$Ci/cm, $P_{des}(Rn)=0.45$, $P_{des}^{eff}(Pb)=0.55$. Regardless of the model parameters, taking into account the finite diameter of the seed with DART2D leads to significantly higher doses compared to the line source 0D approximation.}
    \label{fig:grid doses comparisons}
\end{figure}

Figure \ref{fig:grid doses comparisons} shows the dose calculated for a hexagonal seed lattice of parallel seeds with a grid spacing of 4 mm. As before, the seed radius is 0.35\;mm, its $^{224}$Ra activity is 3\;$\mu$Ci (over 1\;cm length), $P_{des}(Rn)=0.45$, $P_{des}^{eff}(Pb)=0.55$. The calculations are for the low-diffusion high-leakage and high-diffusion low-leakage cases defined above.
The calculation is done for both the full 2D solution and the 0D line-source approximation. The dose at the mid point between three adjacent seeds is 75 / 15\;Gy for the accurate 2D calculation (high- / low-diffusion, respectively) and 57 / 8\;Gy for the line source approximation, emphasizing the need to consider the finite diameter of the seed.  

\section{Summary and conclusions}

In this work we provided closed-form approximations and numerical finite-element schemes for calculating the alpha particle dose of DaRT seeds of finite diameter and length, expanding the discussion of a previous publication \cite{Arazi2020} which modeled seeds as idealized line sources comprised of a collection of point-like segments. The intention of the detailed description of the numerical schemes was to enable other research groups to develop their own codes and test them against ours.

The full 2D calculation presented here was shown to better describe the alpha particle dose field than the point source-based approach, considering the effect of finite seed diameter and the dose fall-off along the seed axis. We therefore recommend using it, rather than the simpler approximation, as the basis for preparing dose lookup tables for DaRT treatment planning.  

The intermediate derivation of closed-form approximations and a finite-element scheme for an infinite cylindrical source can serve two purposes. First, since the full 1D solution (``DART1D'') coincides to high accuracy with the 2D calculation on the seed mid plane, it can be used to validate it. In this respect, it is important to note that due to the rapid buildup of the $^{220}$Rn activity towards its asymptotic form, the accuracy of the closed-form 0D approximation to the $^{220}$Rn+$^{216}$Po alpha dose is better than $10^{-3}$. It can therefore be used for validating the calculation of this part of the dose in both the 1D and 2D finite element schemes. 

Second, given that the full 2D calculation is computationally intensive and can take a long time to complete, the accuracy of DART1D in representing the dose at the seed mid plane allows using it safely for parameter scans to understand the properties of DaRT seed lattices. Here it is important to note that deviations from the mid plane dose profile occur only within $\sim1.5$\;mm from the seed end, as evident from figure \ref{fig:fig5_2DTotalDose_and_radial_axial_doses}. 

The main aim of the text was to outline the numerical scheme of the 1D and 2D solutions, rather than develop computer codes with optimized performance. The most demanding aspect of the calculation is the inversion of the coefficient matrix, which can be optimized and coded more efficiently than presented here. For example, when implemented in Fortran, DART1D completes in $\sim$6\;s a calculation which requires $\sim3$\;m to complete in MATLAB.

It is important to emphasize that while the DL model provides a pragmatic description of the DaRT alpha dose, it is by no means a complete theory. In particular, it does not take into account the possibility of convective effects and nonuniformity in real tumors, which comprise both necrotic and viable regions, with the former evolving as a result of the DaRT treatment itself. However, in spite of its limitations, the diffusion leakage model can provide a quantitative guide for selecting a starting point for treatment planning in clinical trials in terms of seed activity and spacing. This was demonstrated in the first clinical trial \cite{Popovtzer2019}, where 2\;$\mu$Ci seeds were inserted at $\sim5$\;mm spacing, based on the DL model prediction that this would provide a nominal alpha dose of $>10$\;Gy throughout the treated volume, and where $\sim80\%$ of the treated tumors displayed a complete response. 

The predictions of the DL model depend critically on the values of its parameters, in particular, the diffusion lengths of $^{220}$Rn and $^{212}$Pb. Extensive experimental work on measuring them in mouse tumors is ongoing and will be covered in separate publications.

\appendix

\section{Derivation of the asymptotic solutions for infinite cylindrical and line sources} \label{Appendix A}

In this appendix we provide a detailed derivation of the asymptotic solutions of the diffusion-leakage model equations for the number densities of $^{220}$Rn, $^{212}$Pb and $^{212}$Bi, for an infinite cylindrical source and infinite line source.

\subsection{$^{220}$Rn}

The $^{220}$Rn diffusion equation for the case of an infinite cylindrical source of radius $R_0$ in a homogeneous and isotropic medium, in cylindrical coordinates, is:

\begin{equation} \label{eq:Rn dif eq inf cyl source app}
    \frac{\partial n_{Rn}}{\partial t}=\frac{D_{Rn}}{r}\frac{\partial}{\partial r}\left(r\frac{\partial n_{Rn}}{\partial r}\right)+s_{Rn}-\lambda_{Rn}n_{Rn} 
\end{equation}

Assuming no $^{224}$Ra release from the source $s_{Rn}(r,t)=0$, and $^{220}$Rn enters the tumor by direct release from the source surface, with the following boundary condition at $r\rightarrow R_0$:

\begin{equation} \label{eq:Rn_bnd_cond_src_app}
    \lim_{r\rightarrow R_0} 2\pi r j_{Rn}(r,t)=P_{des}(Rn)\frac{\Gamma_{Ra}^{src}(0)}{l} e^{-\lambda_{Ra}t}   
\end{equation}
where $\Gamma_{Ra}^{src}(0)/l$ is the initial $^{224}$Ra activity per unit length of the source, and $j_{Rn}=-D_{Rn}\frac{\partial n_{Rn}}{\partial r}$. Substituting the asymptotic form $n_{Rn}^{asy}(r,t)=\widetilde{n}_{Rn}(r)e^{-\lambda_{Ra}t}$ in eq. (\ref{eq:Rn dif eq inf cyl source app}) leads to:

\begin{equation} \label{eq:n_Rn tilde dif eq inf cyl src app}
    \frac{d^{2}\widetilde{n}_{Rn}}{dr^2}+\frac{1}{r}\frac{d\widetilde{n}_{Rn}}{dr}-\frac{1}{L_{Rn}^2}\widetilde{n}_{Rn}=0 
\end{equation}
where $L_{Rn}$ is defined in eq. (\ref{eq:LRn}). Defining $\xi=r/L_{Rn}$ gives:

\begin{equation}
    \xi^2\frac{d^2\widetilde{n}_{Rn}}{d\xi^2}+\xi\frac{d\widetilde{n}_{Rn}}{d\xi}-\left(\xi^2+n^2\right)\widetilde{n}_{Rn}=0 \qquad n=0
\end{equation}

This has the form of a modified Bessel equation \cite{Abramowitz1972}, for which the solution is:

\begin{equation}
    \widetilde{n}_{Rn}(\xi)=A_{Rn}K_{0}(\xi)+B_{Rn}I_{0}(\xi)
\end{equation}

$I_{0}(\xi)$ is a modified Bessel function of the first kind. Since it diverges for $\xi\rightarrow\infty$, we must have $B_{Rn}=0$. $K_{0}(\xi)$ is a modified Bessel function of the second kind, for which:

\begin{equation}
    K_{0}(\xi)=\int_{0}^{\infty}\frac{\cos\left(\xi t\right)}{\sqrt{t^2+1}}dt
\end{equation}
It vanishes in the limit $\xi\rightarrow\infty$ and hence:

\begin{equation}
    \widetilde{n}_{Rn}(\xi)=A_{Rn}K_{0}(\xi)
\end{equation}
The modified Bessel functions of the second kind have the property that:

\begin{equation} \label{eq:recursion relation Kn derivative app}
    \xi\frac{dK_{n}}{d\xi}=nK_{n}(\xi)-\xi K_{n+1}(\xi) 
\end{equation}
Thus:

\begin{equation}
    \frac{dK_{0}}{d\xi}=-K_{1}(\xi)
\end{equation}
Using this, the current can be expressed as:

\begin{equation} \label{eq: j_Rn cyl app}
    j_{Rn}(r,t)=\frac{D_{Rn}}{L_{Rn}}A_{Rn}K_1\left(\frac{r}{L_{Rn}}\right)e^{-\lambda_{Ra}t} 
\end{equation}
Substituting (\ref{eq: j_Rn cyl app}) into the boundary condition (\ref{eq:Rn_bnd_cond_src_app}) yields:

\begin{equation} \label{eq:A_Rn cyl app}
    A_{Rn}=\frac{P_{des}(Rn)\left(\Gamma_{Ra}^{src}(0)/l\right)}{2\pi D_{Rn} \, (R_0/L_{Rn}) K_1\left(R_0/L_{Rn}\right) } 
\end{equation}
Finally:

\begin{equation} \label{eq:n_Rn asy cyl source final app}
    n_{Rn}^{asy}(r,t)=\frac{P_{des}(Rn)\left(\Gamma_{Ra}^{src}(0)/l\right)}{2\pi D_{Rn} \, (R_0/L_{Rn}) K_1\left(R_0/L_{Rn}\right)}
    K_0\left(\frac{r}{L_{Rn}}\right)e^{-\lambda_{Ra}t} 
\end{equation}
For $\xi\rightarrow0$ \cite{Abramowitz1972}:
\begin{equation} \label{eq:lim xK1(x) app}
    \lim_{\xi\rightarrow 0}\,\xi K_1(\xi)=1  
\end{equation}
Therefore, for the case of an ideal line source ($R_0/L_{Rn}\rightarrow0$):
\begin{equation} \label{n_Rn asy line source final}
    n_{Rn}^{asy}(r,t)=\frac{P_{des}(Rn)\left(\Gamma_{Ra}^{src}(0)/l\right)}{2\pi D_{Rn}} K_0\left(\frac{r}{L_{Rn}}\right)e^{-\lambda_{Ra}t}
\end{equation}

\subsection{$^{212}$Pb}

The diffusion-leakage equation for $^{212}$Pb for the infinite cylindrical source geometry is:

\begin{equation} \label{eq: Pb dif eq cyl app}
    \frac{\partial n_{Pb}}{\partial t}=\frac{D_{Pb}}{r}\frac{\partial}{\partial r}\left(r\frac{\partial n_{Pb}}{\partial r}\right)-\lambda_{Pb}n_{Pb}-\alpha_{Pb}n_{Pb}+\lambda_{Rn}n_{Rn} 
\end{equation}
with the boundary condition:

\begin{equation} \label{eq: n_Pb bnd con cyl src app}
    \lim_{r\rightarrow R_0} 2\pi r j_{Pb}(r,t)=\left(P_{des}^{eff}(Pb)-P_{des}(Rn)\right)\frac{\Gamma_{Ra}^{src}(0)}{l} e^{-\lambda_{Ra}t} 
\end{equation}
where $j_{Pb}=-D_{Pb}\frac{\partial n_{Pb}}{\partial r}$. As before, we are looking for a solution of the form:
\begin{equation} \label{eq:n_Pb asy cyl general form app}
    n_{Pb}^{asy}\left(r,t\right)=\widetilde{n}_{Pb}\left(r\right)e^{-\lambda_{Ra}t} 
\end{equation}

Substituting the asymptotic forms for $^{212}$Pb and $^{220}$Rn in eq. (\ref{eq: Pb dif eq cyl app}) yields:

\begin{equation}
    \fl \qquad \left(\lambda_{Pb}+\alpha_{Pb}-\lambda_{Ra}\right)\widetilde{n}_{Pb}=D_{Pb}\left(\frac{d^{2}\widetilde{n}_{Pb}}{dr^{2}}+\frac{1}{r}\frac{d\widetilde{n}_{Pb}}{dr}\right)+\lambda_{Rn}A_{Rn}K_{0}\left(\frac{r}{L_{Rn}}\right)
\end{equation}
where $A_{Rn}$ is given in (\ref{eq:A_Rn cyl app}). Using the definition of $L_{Pb}$ (\ref{eq:LPb}), this can be written as:

\begin{equation}
    \left(\frac{d^{2}\widetilde{n}_{Pb}}{dr^{2}}+\frac{1}{r}\frac{d\widetilde{n}_{Pb}}{dr}\right)-\frac{1}{L_{Pb}^{2}}\widetilde{n}_{Pb}+\frac{\lambda_{Rn}}{D_{Pb}}A_{Rn}K_{0}\left(\frac{r}{L_{Rn}}\right)=0 \label{eq:nPb_tilda eq cyl source}
\end{equation}
We attempt a solution of the form:

\begin{equation}
    \widetilde{n}_{Pb}\left(r\right)=A_{Pb}K_{0}\left(\frac{r}{L_{Rn}}\right)+B_{Pb}K_{0}\left(\frac{r}{L_{Pb}}\right)
\end{equation}

In order to proceed, we need to utilize some properties of the modified Bessel functions. Again, we define $\xi=r/L$. Using the recursion relation (\ref{eq:recursion relation Kn derivative app}), we have:

\begin{eqnarray}
    \frac{dK_{0}}{d\xi}=-K_{1}\left(\xi\right) \label{eq:dK0_dx} \\
    \frac{dK_{1}}{d\xi}=\frac{1}{\xi}K_{1}\left(\xi\right)-K_{2}\left(\xi\right) \label{eq:dK1_dx}
\end{eqnarray}
Another property of the modified Bessel functions \cite{Abramowitz1972} is:

\begin{equation}
    K_{n+1}\left(\xi\right)=K_{n-1}\left(\xi\right)+\frac{2n}{\xi}K_{n}\left(\xi\right)
\end{equation}
Thus:

\begin{equation}
    K_{2}\left(\xi\right)=K_{0}\left(\xi\right)+\frac{2}{\xi}K_{1}\left(\xi\right) \label{eq:K2=K0+2/xK1}
\end{equation}
Substituting $K_{2}\left(\xi\right)$ from (\ref{eq:K2=K0+2/xK1}) in (\ref{eq:dK1_dx}) gives:

\begin{equation}
    \frac{dK_{1}}{d\xi}=-\left(K_{0}\left(\xi\right)+\frac{1}{\xi}K_{1}\left(\xi\right)\right) \label{eq:dK1_dx final}
\end{equation}
From (\ref{eq:dK0_dx}) we have:
\begin{equation}
    \frac{d}{dr}K_{0}\left(\frac{r}{L}\right) = \frac{1}{L}\frac{d}{d\xi}K_0(\xi) = -\frac{1}{L}K_{1}\left(\frac{r}{L}\right) \label{eq:dK0_dr}
\end{equation}
and from (\ref{eq:dK1_dx final}):
\begin{equation}
    \frac{d}{dr}K_{1}\left(\frac{r}{L}\right)=\frac{1}{L}\frac{d}{d\xi}K_{1}\left(\xi\right)=-\frac{1}{L}\left(K_{0}\left(\frac{r}{L}\right)+\frac{L}{r}K_{1}\left(\frac{r}{L}\right)\right) \label{eq:dK1_dr}
\end{equation}
Using results (\ref{eq:dK0_dr}) and (\ref{eq:dK1_dr}), we get:

\begin{equation}
    \frac{d\widetilde{n}_{Pb}}{dr}=-\frac{A_{Pb}}{L_{Rn}}K_{1}\left(\frac{r}{L_{Rn}}\right)-\frac{B_{Pb}}{L_{Pb}}K_{1}\left(\frac{r}{L_{Pb}}\right) \label{eq:dn_Pb_tilda/dr}
\end{equation}
and:

\begin{eqnarray}
    \fl \quad \frac{d^{2}\widetilde{n}_{Pb}}{dr^{2}}= \frac{A_{Pb}}{L_{Rn}^{2}}\left(K_{0}\left(\frac{r}{L_{Rn}}\right)+\frac{L_{Rn}}{r}K_{1}\left(\frac{r}{L_{Rn}}\right)\right) +\frac{B_{Pb}}{L_{Pb}^{2}}\left(K_{0}\left(\frac{r}{L_{Pb}}\right)+\frac{L_{Pb}}{r}K_{1}\left(\frac{r}{L_{Pb}}\right)\right) \nonumber \\
\end{eqnarray}
resulting in:

\begin{equation}
    \frac{d^{2}\widetilde{n}_{Pb}}{dr^{2}}+\frac{1}{r}\frac{d\widetilde{n}_{Pb}}{dr}=\frac{A_{Pb}}{L_{Rn}^{2}}K_{0}\left(\frac{r}{L_{Rn}}\right)+\frac{B_{Pb}}{L_{Pb}^{2}}K_{0}\left(\frac{r}{L_{Pb}}\right) \label{eq:d2nPb/dr2 + 1/r dnPb/dr}
\end{equation}
When (\ref{eq:d2nPb/dr2 + 1/r dnPb/dr}) is inserted in eq. (\ref{eq:nPb_tilda eq cyl source}), the $K_{0}\left(r/L_{Pb}\right)$ terms cancel out, leaving:

\begin{equation} \label{eq:A_Pb cyl source app}
    \fl \quad A_{Pb}=\frac{L_{Rn}^{2}L_{Pb}^{2}}{L_{Rn}^{2}-L_{Pb}^{2}}\frac{\lambda_{Rn}}{D_{Pb}}A_{Rn}=\frac{L_{Rn}^{2}L_{Pb}^{2}}{L_{Rn}^{2}-L_{Pb}^{2}} \, \frac{\lambda_{Rn}}{D_{Pb}} \, \frac{P_{des}(Rn)\left(\Gamma_{Ra}^{src}(0)/l\right)}{2\pi D_{Rn}\, (R_0/L_{Rn})K_1\left(R_0/L_{Rn}\right) } 
\end{equation}

The coefficient $B_{Pb}$ is found from the boundary condition (\ref{eq: n_Pb bnd con cyl src app}). Defining $\widetilde{j}_{Pb}\left(r\right)=-D_{Pb}\frac{d}{dr}\widetilde{n}_{Pb}\left(r\right)$ and using (\ref{eq:dn_Pb_tilda/dr}), we have:

\begin{eqnarray}
    \fl \qquad 2\pi R_0 \widetilde{j}_{Pb}(R_0) & = 2\pi R_0 D_{Pb} \left( \frac{A_{Pb}}{L_{Rn}} K_{1}\left(\frac{R_0}{L_{Rn}}\right) + \frac{B_{Pb}}{L_{Pb}} K_{1}\left(\frac{R_0}{L_{Pb}}\right) \right) \nonumber \\
    & = \left(P_{des}^{eff}(Pb)-P_{des}(Rn)\right)\frac{\Gamma_{Ra}^{src}(0)}{l}
\end{eqnarray}
yielding:

\begin{equation} \label{eq:B_Pb cyl source app}
    \fl \qquad B_{Pb} = \frac{\left(P_{des}^{eff}(Pb)-P_{des}(Rn)\right)\left(\Gamma_{Ra}^{src}(0)/l\right)}{ 2\pi D_{Pb} \, (R_0/L_{Pb})K_1(R_0/L_{Pb}) } - A_{Pb}\frac{ (R_0/L_{Rn})K_1(R_0/L_{Rn}) }{ (R_0/L_{Pb})K_1(R_0/L_{Pb}) } 
\end{equation}
For an ideal line source, $(R_0/L)K_1(R_0/L) \rightarrow1 $ giving:

\begin{equation} \label{eq:A_Pb line source}
    A_{Pb}^{line} = \frac{L_{Rn}^{2}L_{Pb}^{2}}{L_{Rn}^{2}-L_{Pb}^{2}} \frac{\lambda_{Rn}}{D_{Pb}} \frac{P_{des}(Rn)\left(\Gamma_{Ra}^{src}(0)/l\right)}{2\pi D_{Rn} } 
\end{equation}

\begin{equation} \label{eq:B_Pb line source}
    B_{Pb}^{line} = \frac{\left(P_{des}^{eff}(Pb)-P_{des}(Rn)\right)\left(\Gamma_{Ra}^{src}(0)/l\right)}{ 2\pi D_{Pb} } - A_{Pb} 
\end{equation}
Finally, we have:

\begin{equation} \label{eq:n_Pb_asy cyl source final result app}
    n_{Pb}^{asy}(r,t)=\left(A_{Pb}K_{0}\left(\frac{r}{L_{Rn}}\right)+B_{Pb}K_{0}\left(\frac{r}{L_{Pb}}\right)\right)e^{-\lambda_{Ra}t} 
\end{equation}

\subsection{$^{212}$Bi}

The diffusion-leakage equation for $^{212}$Bi for an infinite cylindrical source is:

\begin{equation} \label{eq:Bi212 dif eq cyl source app}
    \frac{\partial n_{Bi}}{\partial t}=\frac{D_{Bi}}{r}\frac{\partial}{\partial r}\left(r\frac{\partial n_{Bi}}{\partial r}\right)-\lambda_{Bi}n_{Bi}-\alpha_{Bi}n_{Bi}+\lambda_{Pb}n_{Pb} 
\end{equation}
Since $^{212}$Bi atoms are not emitted directly from the source, the boundary condition is:

\begin{equation} \label{eq:Bi212 bnd cond cyl source app}
    \lim_{r\rightarrow R_0}2\pi r j_{Bi}(r,t)=0 
\end{equation}
We look for an asymptotic solution of the form: $n_{Bi}^{asy}(r,t)=\widetilde{n}_{Bi}(r)e^{-\lambda_{Ra}t}$, attempting:

\begin{equation} \label{eq:n_Bi_tilda general form cyl source}
    \widetilde{n}_{Bi}(r)=A_{Bi}K_{0}\left(\frac{r}{L_{Rn}}\right)+B_{Bi}K_{0}\left(\frac{r}{L_{Pb}}\right)+C_{Bi}K_{0}\left(\frac{r}{L_{Bi}}\right) 
\end{equation}

Substituting the asymptotic form (\ref{eq:n_Bi_tilda general form cyl source}) in eq. (\ref{eq:Bi212 dif eq cyl source app}) and in the boundary condition (\ref{eq:Bi212 bnd cond cyl source app}), using the asymptotic form of the $^{212}$Pb solution (\ref{eq:n_Pb_asy cyl source final result app}) and the expression for the effective $^{212}$Bi diffusion length (\ref{eq:LBi}), gives:

\begin{equation}
\fl \qquad n_{Bi}^{asy}(r,t)=\left(A_{Bi}K_{0}\left(\frac{r}{L_{Rn}}\right)+B_{Bi}K_{0}\left(\frac{r}{L_{Pb}}\right)+C_{Bi}K_{0}\left(\frac{r}{L_{Bi}}\right)\right)e^{-\lambda_{Ra}t}
\end{equation}
For a cylindrical source of radius $R_0$:

\begin{equation} \label{eq:A_Bi cyl source app}
    A_{Bi}=\frac{L_{Rn}^{2}L_{Bi}^{2}}{L_{Rn}^{2}-L_{Bi}^{2}}\frac{\lambda_{Pb}}{D_{Bi}}A_{Pb} 
\end{equation}

\begin{equation} \label{eq:B_Bi cyl source app}
    B_{Bi}=\frac{L_{Pb}^{2}L_{Bi}^{2}}{L_{Pb}^{2}-L_{Bi}^{2}}\frac{\lambda_{Pb}}{D_{Bi}}B_{Pb} 
\end{equation}

\begin{equation} \label{eq: C_Bi cyl source app}  
    C_{Bi}=- \frac{ (R_0/L_{Rn})K_1(R_0/L_{Rn})A_{Bi} + (R_0/L_{Pb})K_1(R_0/L_{Pb})B_{Bi} }{ (R_0/L_{Bi})K_1(R_0/L_{Bi}) }
\end{equation}
where $A_{Pb}$ and $B_{Pb}$ are given in (\ref{eq:A_Pb cyl source app}) and (\ref{eq:B_Pb cyl source app}). For an infinite line source, $A_{Pb}$ and $B_{Pb}$ are replaced by their line source forms (\ref{eq:A_Pb line source}) and (\ref{eq:B_Pb line source}) and:

\begin{equation}
    C_{Bi}^{line}=-(A_{Bi}+B_{Bi})
\end{equation}

\section{Discretization of the DL model equations in cylindrical coordinates at the seed wall} \label{Appendix B}

The discretization of the DL equations for elements (cylindrical shells or rings) touching the seed wall requires special attention, making use of the appropriate boundary conditions. We demonstrate this in 1D, with practically the same procedure for the 2D case. As stated in the main text, the radial coordinate of the $i$-th cylindrical shell is given for the shell center: $r_i=R_0+(i-\frac{1}{2})\Delta r$. For $i>1$ the diffusive term in the 1D DL equations is discretized as follows:

\begin{eqnarray}
    \fl \qquad \frac{D}{r}\frac{\partial}{\partial r}\left(r\frac{\partial n}{\partial r}\right)_{r_i} \Longrightarrow \nonumber \\
    \fl \qquad \frac{D}{r_i}\frac{\left(r\frac{\partial n}{\partial r}\right)_{r_i+\Delta r/2}-\left(r\frac{\partial n}{\partial r}\right)_{r_i-\Delta r/2}}{\Delta r} =
    \frac{D}{r_i} \frac{(r_i+\frac{\Delta r}{2})\frac{n_{i+1}-n_i}{\Delta r}-(r_i-\frac{\Delta r}{2})\frac{n_{i}-n_{i-1}}{\Delta r}}{\Delta r} = \nonumber \\
    \fl \qquad \frac{D}{\Delta r^2}\left( \left(1+\frac{\Delta r}{2r_i}\right)n_{i+1} + \left(1-\frac{\Delta r}{2r_i}\right)n_{i-1} - 2n_i \right)
\end{eqnarray}

\noindent{In the special case $i=1$, i.e., the shell immediately outside the source surface with $r_1=R_0+\Delta r/2$, we use the boundary conditions at $r\rightarrow R_0$. Since the current density is $j=-D\left(\partial n/\partial r\right)$, we can write}:

\begin{eqnarray}
    \fl \qquad \frac{D}{r}\frac{\partial}{\partial r}\left(r\frac{\partial n}{\partial r}\right)_{r_1} \Longrightarrow \nonumber \\
    \fl \qquad \frac{D}{R_0+\Delta r/2}\frac{\left(R_0+\Delta r\right) \left(\frac{\partial n}{\partial r}\right)_{R_0+\Delta r}-R_0\left(\frac{\partial n}{\partial r}\right)_{R_0}}{\Delta r} = 
    \frac{D}{R_0+\Delta r/2} \frac{\left(R_0+\Delta r\right)\frac{n_2-n_1}{\Delta r}+R_0\frac{j(R_0,t)}{D}}{\Delta r} \nonumber \\
\end{eqnarray}

\noindent{The source boundary conditions eq. (\ref{eq:Rn_bnd_cond_src})-(\ref{eq:Bi_bnd_cond_src}) give:}

\begin{eqnarray}
    R_0 j_{Rn}(R_0,t) = \frac{1}{2\pi}P_{des}(Rn)\frac{\Gamma_{Ra}^{src}(0)}{l}e^{-\lambda_{Ra}t} \\
    R_0 j_{Pb}(R_0,t) = \frac{1}{2\pi}\left(P_{des}^{eff}(Pb)-P_{des}(Rn))\right)\frac{\Gamma_{Ra}^{src}(0)}{l}e^{-\lambda_{Ra}t} \\
    R_0 j_{Bi}(R_0,t) = 0
\end{eqnarray}
Therefore:
\bigskip
\begin{eqnarray}
    \fl \qquad \frac{D_{Rn}}{r}\frac{\partial}{\partial r}\left(r\frac{\partial n_{Rn}}{\partial r}\right)_{r_1} \Longrightarrow \nonumber \\ 
    \fl \qquad \frac{D_{Rn}}{\Delta r^2}\left(\frac{1+\Delta r/R_0}{1+\Delta r/2R_0}\right)\left(n_{Rn,2}-n_{Rn,1}\right) + \frac{P_{des}(Rn)\left(\Gamma_{Ra}^{src}(0)/l\right) e^{-\lambda_{Ra}t}}{2\pi R_0\Delta r\left(1+\Delta r/2R_0\right)} \nonumber \\
\end{eqnarray}

\begin{eqnarray}
    \fl \qquad \frac{D_{Pb}}{r}\frac{\partial}{\partial r}\left(r\frac{\partial n_{Pb}}{\partial r}\right)_{r_1} \Longrightarrow \nonumber \\
    \fl \qquad \frac{D_{Pb}}{\Delta r^2}\left(\frac{1+\Delta r/R_0}{1+\Delta r/2R_0}\right)\left(n_{Pb,2}-n_{Pb,1}\right) + \frac{\left(P_{des}^{eff}(Pb)-P_{des}(Rn)\right)\left(\Gamma_{Ra}^{src}(0)/l\right) e^{-\lambda_{Ra}t}}{2\pi R_0\Delta r\left(1+\Delta r/2R_0\right)} \nonumber \\
\end{eqnarray}
and
\begin{eqnarray}
    \fl \qquad \frac{D_{Bi}}{r}\frac{\partial}{\partial r}\left(r\frac{\partial n_{Bi}}{\partial r}\right)_{r_1} \Longrightarrow 
    \frac{D_{Bi}}{\Delta r^2}\left(\frac{1+\Delta r/R_0}{1+\Delta r/2R_0}\right)\left(n_{Bi,2}-n_{Bi,1}\right)
\end{eqnarray}

\section*{References}

\bibliography{DaRT_finite_element_modeling_bib}

\end{document}